\documentclass[a4paper,11pt]{article}

\pdfoutput=1

\usepackage{amsmath,amssymb,mathtools}
\usepackage{color}
\usepackage{graphicx}
\usepackage{subfigure}
\usepackage{cite}
\usepackage{hyperref}
\usepackage{multirow,makecell}
\usepackage{textcomp}
\usepackage{wasysym}

\usepackage{amsthm}

\usepackage{ulem} 

\theoremstyle{definition}

\usepackage[text={17cm,24.6cm},centering]{geometry} 

\numberwithin{equation}{section}

\newcommand{\mlr}[2]{\multirow{#1}*{#2}} 
\newcommand{\mlc}[3]{\multicolumn{#1}{#2}{#3}} 

\newcommand{\rot}[2]{\multirow{#1}*{\rotatebox[origin=cc]{90}{#2}}} 

\def \be {\begin{equation}}
\def \ee {\end{equation}}
\def \ba {\begin{array}}
\def \ea {\end{array}}
\def \bea{\begin{eqnarray}}
\def \eea{\end{eqnarray}}
\def \nn {\nonumber}

\def \a {\alpha}
\def \b {\beta}
\def \g {\gamma}

\def \G {\Gamma}
\def \d {\delta}

\def \e {\epsilon}
\def \ve {\varepsilon}

\def \l {\lambda}

\def \s {\sigma}

\def \r {\rho}

\def \O {\Omega}

\def \t {\tau}

\def \mA {\mathcal A}
\def \mB {\mathcal B}

\def \mD {\mathcal D}
\def \mE {\mathcal E}

\def \mH {\mathcal H}
\def \mI {\mathcal I}

\def \mO {\mathcal O}

\def \mR {\mathcal R}

\def \mX {\mathcal X}
\def \mY {\mathcal Y}

\def \cA {{\mathcal A}}
\def \cB {{\mathcal B}}

\def \cD {{\mathcal D}}

\def \cO {{\mathcal O}}

\def \rC {{\mathrm C}}

\def \rI {{\mathrm I}}

\def \p {\partial}

\def \f {\frac}

\def \mc {\mathcal}

\def \sr {\sqrt}
\def \td {\tilde}

\def \pp {\propto}
\def \inf {\infty}

\def \lag {\langle}
\def \rag {\rangle}

\def \ep {\mathrm{e}}
\def \ii {\mathrm{i}}

\def \tr {\textrm{tr}}

\def \and {{~\textrm{and}~}}

\def \CFT {{\textrm{CFT}}}

\def \Trho {\langle T \rangle_\rho}
\def \Arho {\langle \mathcal A \rangle_\rho}
\def \Brho {\langle \mathcal B \rangle_\rho}
\def \Drho {\langle \mathcal D \rangle_\rho}

\def \Tsig {\langle T \rangle_\sigma}
\def \Asig {\langle \mathcal A \rangle_\sigma}
\def \Bsig {\langle \mathcal B \rangle_\sigma}
\def \Dsig {\langle \mathcal D \rangle_\sigma}

\begin{document}

\title{\textbf{Note on ETH of descendant states in 2D CFT}}
\author{
Wu-zhong Guo$^{1}$\footnote{wzguo@cts.nthu.edu.tw}~,
Feng-Li Lin$^{2}$\footnote{fengli.lin@gmail.com}~
and
Jiaju Zhang$^{3,4}$\footnote{jiaju.zhang@unimib.it}
}
\date{}

\maketitle
\vspace{-10mm}
\begin{center}
{\it
$^{1}$Physics Division, National Center for Theoretical Sciences, National Tsing Hua University,\\
No.\ 101, Sec.\ 2, Kuang Fu Road, Hsinchu 30013, Taiwan\\\vspace{1mm}
$^{2}$Department of Physics, National Taiwan Normal University,\\
No.\ 88, Sec.\ 4, Ting-Chou Road, Taipei 11677, Taiwan\\\vspace{1mm}
$^{3}$Dipartimento di Fisica G.\ Occhialini, Universit\`a degli Studi di Milano-Bicocca,\\Piazza della Scienza 3, 20126 Milano, Italy\\\vspace{1mm}
$^{4}$SISSA and INFN, Via Bonomea 265, 34136 Trieste, Italy
}
\vspace{10mm}
\end{center}

\begin{abstract}

  We investigate the eigenstate thermalization hypothesis (ETH) of highly excited descendant states in two-dimensional large central charge $c$ conformal field theory. We use operator product expansion of twist operators to calculate the short interval expansions of entanglement entropy and relative entropy for an interval of length $\ell$ up to order $\ell^{12}$. Using these results to ensure ETH of a heavy state when compared with the canonical ensemble state up to various orders of $c$, we get the constraints on the expectation values of the first few quasiprimary operators in the vacuum conformal family at the corresponding order of $c$. Similarly, we also obtain the constraints from the expectation values of the first few Korteweg-de Vries charges. We check these constraints for some types of special descendant excited states. Among the descendant states we consider, we find that at most only the leading order ones of the ETH constraints can be satisfied for the descendant states that are slightly excited on top of a heavy primary state. Otherwise, the ETH constraints are violated for the descendant states that are heavily excited on top of a primary state.

\end{abstract}

\baselineskip 18pt
\thispagestyle{empty}
\newpage

\tableofcontents


\section{Introduction}

Eigenstate thermalization hypothesis (ETH) \cite{Deutsch:1991,Srednicki:1994,Srednicki:1995pt} states that a typical highly excited energy eigenstate in a quantum chaotic system behaves like a thermal state.
It explains how various statistical ensembles emerge in quantum many-body systems, and forms the cornerstone for the studies of quantum thermalization.
ETH was originally formulated in terms of local operators in the basis of energy eigenstates and can be called local ETH.
When local ETH applies, the expectation values of a general local operator $\mO$ in the basis of high energy eigenstates $\{ |a\rag \}$ with $\lag a | b \rag=\d_{ab}$ take the form
\be\label{localETH}
\lag a | \mO | b \rag = \mO(E)\d_{ab} + \ep^{-O(S(E))},
\ee
where $\mO(E)$ is a smooth function of $E=(E_a+E_b)/2$, and $S(E)$ is the entropy of the microcanonical ensemble state with energy $E$.

As a generalization of local ETH, subsystem ETH was proposed in \cite{Lashkari:2016vgj,Dymarsky:2016aqv} and one compares the reduced density matrix (RDM) of the excited energy eigenstate with the RDM of the corresponding thermal state. For a system of size $V$, one chooses a subsystem $A$ with size $V_A$, and its complement can be denoted as $\bar A$.
The RDM $\r_A$ of the subsystem $A$ in a state $\r$ of the total system is defined by tracing out the degrees of freedom of $\bar A$, i.e., $\r_A := \tr_{\bar A}\r$. For a high energy eigenstate $|a\rag$, subsystem ETH is defined in the thermodynamic limit $V \to \inf$ with $V_A / V \to 0$ and $E_a/V$ being finite. When subsystem ETH applies, the RDM of a high energy eigenstate $\r_{A,a}=\tr_{\bar A}|a\rag\lag a|$ is {sufficiently} close to a universal RDM $\r_{A,\rm{ETH}}(E)$, {that is, their trace distance
\be\label{subETH}
\| \r_{A,a} - \r_{A,\rm{ETH}}(E) \| \sim \ep^{-O(S(E))},
\ee
where $E=E_a$ and $S(E)$ is thermal entropy of the microcanonical ensemble state at energy $E$.} The trace distance of {density matrices $\r$ and $\s$ is} defined by $\|\r-\s\|:=\f12\sum_i|\l_i|$ with {$\l_i$'s} being the eigenvalues of the hermitian matrix $\r-\s$. The universal RDM $\r_{A,\rm{ETH}}(E)$ is {sufficiently} close to some specific ensemble {in the same sense of (\ref{subETH}).}

{The above statements of ETH are succinct, however, the explicit check by evaluating the trace distance for particular state is usually technically involved and numerical implementation is needed. It is then interesting to examine ETH with some other quantities, and at the same time to consider the systems for which these quantities can be evaluated analytically.   One of such systems is the two-dimensional (2D) conformal field theory (CFT) with large central charge $c$, which is} dual to quantum gravity in three-dimensional (3D) anti-de Sitter (AdS) space with a small Newton constant $G\sim 1/c$ \cite{Brown:1986nw}. This duality  is the precursor of AdS/CFT correspondence \cite{Maldacena:1997re,Gubser:1998bc,Witten:1998qj}. Moreover, the thermal CFT state is dual to black hole in 3D AdS space, i.e. the Ba\~nados-Teitelboim-Zanelli (BTZ) black hole \cite{Banados:1992wn}. Thus the study of ETH in 2D large $c$ CFT could shed light on the nature of black hole microstates and the information loss paradox.
In fact, both the diagonal and off-diagonal parts of the local ETH in 2D CFT {are} consistent with the coarse-grained results obtained from modular covariance of the one-point and multi-points functions on a torus \cite{Kraus:2016nwo,Brehm:2018ipf,Romero-Bermudez:2018dim,Hikida:2018khg}.

{Even though the trace distance is a canonical measure to distinguish two RDMs in the context of subsystem ETH, it is in general difficult to calculate in quantum field theory such as CFT. Instead, we can consider some other quantities for the same purpose despite that the results may not be as strong as the ones from trace distance.} The quantities we will consider in this paper include the entanglement entropy (EE), the relative entropy, and the expectation values of Korteweg-de Vries (KdV) charges. The EE of a RDM $\r_A$ is defined as
\be
S_A =  - \tr_A (\r_A \log \r_A).
\ee
The relative entropy of two RDMs $\r_A$ and $\s_A$ is defined as
\be
S(\r_A\|\s_A) = \tr_A (\r_A \log \r_A) - \tr_A (\r_A \log \s_A),
\ee
and it characterizes the difference between these two RDMs. {For the purpose of ETH, we only need to tell the difference between excited state and thermal state {\it locally}. Thus, we can use the operator product expansion (OPE) of twist operators \cite{Calabrese:2004eu,Headrick:2010zt,Calabrese:2010he,Chen:2013kpa} to calculate the {\it short interval} expansion of EE and relative entropy for both excited state and thermal state, and then compare them to examine the subsystem ETH. On the other hand, we can also check the local ETH only for some particular set of local operators, which are the more physically relevant observables such as conserved charges. In 2D CFTs, there are such a set of operators, which are the infinite number of mutually commuting conserved KdV charges $Q_{2k-1}$, $k=1,2,\cdots$ \cite{Sasaki:1987mm,Eguchi:1989hs,Bazhanov:1994ft}.  In the appendix (\ref{KdVgeneral}). we give the expressions of the first few KdV charges.}

By investigating the two-point functions of light operators and the EE, it was found that {at the leading order of large $c$ expansion, the heavy primary states behave like the canonical ensemble state so that ETH is justified} \cite{Fitzpatrick:2014vua,Fitzpatrick:2015zha,Asplund:2014coa,Caputa:2014eta}.
With $1/c$ corrections, it was found that the RDMs of the primary excited state and canonical ensemble state are in fact different {when} comparing their R\'enyi entropy, entanglement entropy, relative entropy, trace square distance, and other quantities \cite{Lashkari:2016vgj,Lin:2016dxa,He:2017vyf,Basu:2017kzo,He:2017txy,Lashkari:2017hwq}. A possible resolution is to replace the canonical ensemble state with the generalized Gibbs ensemble (GGE) state \cite{Rigol:2006} by including the KdV charges and their corresponding chemical potentials \cite{He:2017vyf,Basu:2017kzo,He:2017txy,Lashkari:2017hwq}. We will briefly discuss ETH in the context of GGE in the conclusion part of this paper.

In this paper, we investigate ETH of {some types of descendant} states with conformal weights $h \gg c \gg 1$ in a 2D large $c$ CFT in the thermodynamic limit. Previous studies of ETH in 2D CFT \cite{Fitzpatrick:2014vua,Fitzpatrick:2015zha,Asplund:2014coa,Caputa:2014eta,Lashkari:2016vgj,Lin:2016dxa,He:2017vyf,Basu:2017kzo,He:2017txy,Lashkari:2017hwq} focus only on the primary excited states. {However, the ETH is formulated for the general energy eigenstates, regardless of these states being primary or descendant. Though the properties of descendant states are algebraically determined by those of primary states, it does not necessarily imply that they also satisfy ETH  as the primary states do  even at the leading order of large $c$ expansions. Moreover, there exists an infinite tower of descendant states for each primary state. Thus, it is important to check the ETH for the descendant states to have a full understanding  of quantum thermalization in 2D CFT.}

In this paper we will investigate the subsystem ETH {for three types of special descendant states, depending on how heavily they are excited on top of a primary state.  We find that only descendant states which are slightly excited on top of a heavy primary state can satisfy the ETH at leading order of the large $c$ expansions. Even we only consider some special descendant states, our intuitive results found in this paper may give hints for the further studies on the general patterns of quantum thermalization for the generic eigenstates.  Moreover, from the point of view of AdS/CFT correspondence, the non-thermal descendant states found here cannot be the dual to black hole microstates. This adds more weights to the puzzle of black hole information paradox, i.e. how the thermality of black hole arises.}

The remaining part of the paper is organized as follows.
In section~\ref{seccon} {we examine ETH for a generic state by comparing its EE, relative entropy and expectation values of KdV charges with the counterparts of the corresponding {canonical ensemble state}. From these, we obtain the constraints on the expectation values of the first few quasiprimary operators in the vacuum conformal family with respect to this excited state at different orders of $c$.}
In section~\ref{secche} we check these constraints for the primary states and some various descendant states.
We conclude with discussion in section \ref{seccnd}.
We collect various calculation details in the appendices.
In appendix~\ref{appCFT} we give some useful details of the quasiprimary operators and their correlation functions in 2D CFTs, including both reviews and new calculations.
In appendix~\ref{appexp} we first review the short interval expansions of the EE and relative entropy of an interval of length $\ell$ with the details of enumerating the quasiprimary operators, and then calculate the results up to order $\ell^{12}$, which is higher than the order $\ell^{8}$ in literature. The new results up to order $\ell^{12}$ is crucial to our calculations in the main text.
In appendix~\ref{apppro} we give the proof of a statement used in the previous appendix.

\section{Constraints of expectation values from ETH} \label{seccon}

We consider a 2D large $c$ CFT on a cylinder with spatial period $L$, and choose the subregion $A$ as a short interval of length $\ell$ ($\ll L$). In this setup, we define ETH in terms of various quantities such as EE, relative entropy and expectation values of KdV charges at different orders of $c$. Our method of evaluating these quantities and the results are reviewed and given in the appendices. Using the short interval expansions, in appendix~\ref{appexp} we calculate the EE (\ref{EEgeneral}) and relative entropy (\ref{relativegeneral}) for general translation-invariant states up to order $\ell^{12}$. The first few KdV charges written in terms of quasiprimary operators are listed in (\ref{KdVgeneral}). Based on these results for each quantity, we require ETH to hold for the considered state and extract the associated constraints on the expectations values of the first few quasiprimary operators in vacuum conformal family.

As shown in \cite{Lashkari:2017hwq} {(see also \cite{Hartman:2013mia} for the earlier results on EE and R\'enyi entropy)}, when the subsystem ETH is satisfied the universal RDM is solely expressed in terms of the operators in the vacuum conformal family. {This is why we consider only the constraints in the vacuum conformal family.} On the other hand, in the OPE of twist operators the contributions from the holomorphic and anti-holomorphic sectors factorize, and the contributions from the two sectors are similar. Without loss of generality, we only include the contributions from the holomorphic sector of the vacuum conformal family, and the addition of the anti-holomorphic sector is straightforward. The generalizations of the results to non-chiral states in non-chiral theories are easy.

To define ETH, we also need to take the thermodynamic limit \cite{Lashkari:2016vgj}, i.e. taking the energy $E$ and the total length $L$ to infinity  but keeping the energy density $\ve=E/L$ to be finite. In the thermodynamic limit, the inverse temperature $\b$ and the interval length $\ell$ satisfy $\b/L \to 0$, $\ell/L \to 0$, but there is no requirement for $\ell/\b$.
All the constraints in this section should be understood as under the thermodynamic limit.
To do short interval expansions of the EE and relative entropy we further require $\ell \ll \b$. In summary we need $\ell \ll \b \ll L$, and the constraints should satisfy for all orders of the expansion of $\ell/\b$.

\subsection{Constraints for all orders of large $c$}

We use the EE (\ref{EEgeneral}), relative entropy (\ref{relativegeneral}), and the expectation values of the KdV charges (\ref{KdVgeneral}) to get the constraints for the expectation values of the first few quasiprimary operators $T$, $\cA$, $\cB$, $\cD$ in the holomorphic sector of the vacuum conformal family. Here $T$ is the stress tensor, one can see the definition of $\mA$ in (\ref{Adef}) and the definitions of $\mB$, $\mD$ in (\ref{BDdef}).

\begin{figure}[htbp]
  \centering
  \includegraphics[width=0.3\textwidth]{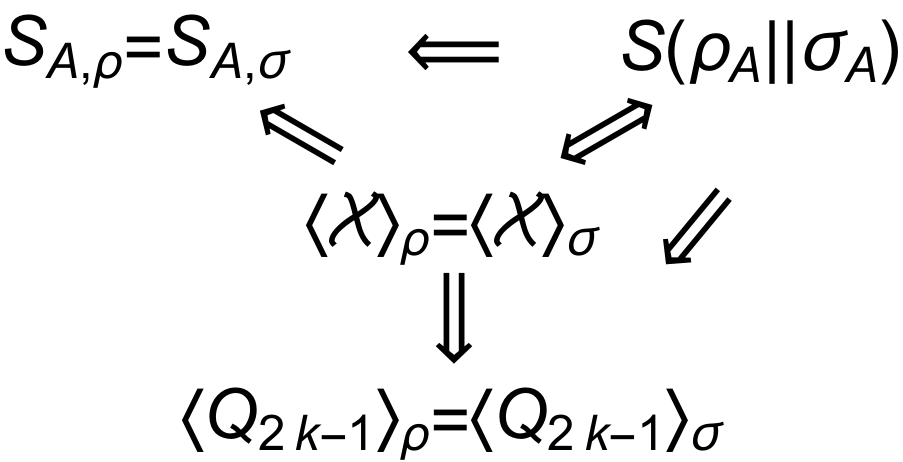}\\
  \caption{Relations of the three sets of constraints from the EE (\ref{genconEE}), relative entropy (\ref{genconRE}) and  expectation values of KdV charges (\ref{genconKdV}).}\label{gencon}
\end{figure}

 By using the formula of EE (\ref{EEgeneral}),
for two general states $\r$, $\s$ of the whole system, requiring $S_{A,\r}=S_{A,\s}$ we get the constraints
\bea \label{genconEE}
\lag T \rag_\r= \lag T \rag_\s, ~~
\lag \mA \rag_\r= \lag \mA \rag_\s ~\mathrm{or}~
\lag \mA \rag_\r + \lag \mA \rag_\s = \f{2(5c+22)}{5c}\lag T \rag_\s^2, ~~
\cdots.
\eea
 By using the formula of relative entropy (\ref{relativegeneral}),
from $S(\r_A\|\s_A)=0$, we get
\be \label{genconRE}
\lag T \rag_\r= \lag T \rag_\s, ~~
\lag \mA \rag_\r= \lag \mA \rag_\s, ~~
\lag \mB \rag_\r= \lag \mB \rag_\s, ~~
\lag \mD \rag_\r= \lag \mD \rag_\s, ~~
\cdots.
\ee
In fact $S(\r_A\|\s_A)=0$ is equivalent to $\r_A=\s_A$, and this leads to $\lag \mX \rag_\r= \lag \mX \rag_\s$ for all local operators $\{\mX\}$.
By using the formulas of KdV charges (\ref{KdVgeneral}),
for the expectation values of KdV charges $\lag Q_{2k-1} \rag_\r = \lag Q_{2k-1} \rag_\s$, $k=1,2,3,\cdots$, we get
\be \label{genconKdV}
\lag T \rag_\r = \lag T \rag_\s, ~~
\lag \mA \rag_\r = \lag \mA \rag_\s, ~~
\lag \mD \rag_\r = \lag \mD \rag_\s  + \frac{25 (2 c+7) (7 c+68)}{108 (70 c+29)} ( \lag \mB \rag_\r - \lag \mB \rag_\s ), ~~
\cdots.
\ee
Generally, the three sets of constraints are not equivalent, and their relations are summarized in figure~\ref{gencon}. In fact, two states having the same expectation values of KdV charges, i.e. satisfying (\ref{genconKdV}), do not necessarily satisfy (\ref{genconRE}) and it can lead to a possible non-vanishing relative entropy
\be \label{SrAsA}
S(\r_A\|\s_A) = \frac{25 ( 5 c^2+203c+791) \ell^{12}}{105080976 c (2 c-1) (5 c+22)} (\lag \mB \rag_\r - \lag \mB \rag_\s)^2 + O(\ell^{14}).
\ee

More specifically for requiring ETH, we choose the state $\sigma$ to be the canonical ensemble state $\r_\b$ and compare its RDM $\r_{A,\b}$ with the RDM $\r_{A}$ of the state $\r$. Because the modular Hamiltonian of RDM $\r_{A,\b}$ is a local integral of the stress tensor \cite{Wong:2013gua},  requiring $S(\r_A\|\r_{A,\b})=0$ or $S(\r_{A,\b}\|\r_A)=0$ is equivalent to the condition $S_A=S_{A,\b}$  \cite{Lashkari:2016vgj}, which then yields for all orders of $c$,
\be \label{conallEERE}
\lag T \rag_\r = \lag T \rag_\b, ~~
\lag \mA \rag_\r = \lag \mA \rag_\b, ~~
\lag \mB \rag_\r = \lag \mB \rag_\b, ~~
\lag \mD \rag_\r = \lag \mD \rag_\b, ~~
\cdots,
\ee
with the expectation values in canonical ensemble state as given (\ref{TABDb}).
One can see the expectation values of KdV charges of the thermal state as given in (\ref{KdVb}).
On the other hand, By requiring that the expectation values of KdV charges $\lag Q_{2k-1} \rag_\r = \lag Q_{2k-1} \rag_\b$, $k=1,2,3,\cdots$, we get
\be \label{conallKdV}
\lag T \rag_\r = \lag T \rag_\b, ~~
\lag \mA \rag_\r = \lag \mA \rag_\b, ~~
\lag \mD \rag_\r = \lag \mD \rag_\b  + \frac{25 (2 c+7) (7 c+68)}{108 (70 c+29)} ( \lag \mB \rag_\r - \lag \mB \rag_\b ), ~~
\cdots.
\ee
Constraints (\ref{conallKdV}) {are not completely the same with (\ref{conallEERE})  so that a state $\r$ satisfying (\ref{conallKdV}) can possibly lead to} non-vanishing relative entropy.

\subsection{Constraints for fixed orders of large $c$}

In the thermodynamical limit, the exact form of the canonical ensemble state EE is known \cite{Calabrese:2004eu}
\be \label{SAb}
S_{A,\b} = \f{c}{6} \log\Big( \f{\b}{\pi\e} \sinh\f{\pi\ell}{\b} \Big).
\ee
We may relax the ETH condition up to different orders of $c$, i.e. by requiring $S_A-S_{A,\b}$ to be at a specific order of $c$.
If we just require $S_A-S_{A,\b}=O(c^0)$, then we get the leading order constraints of EE
\bea \label{conEEL}
&& \lag T \rag_\r = -\frac{\pi^2 c}{6 \beta^2}+{t_0}+O(1/c), ~~
   \lag \mA \rag_\r = \frac{\pi^4 c^2}{36 \beta^4}+ a_1 c + O(c^0), \nn\\
&& \lag \mB \rag_\r = O(c), ~~
   \lag \mD \rag_\r = -\frac{\pi^6 c^3}{216 \beta^6}
                      -\frac{c^2 ( \pi^4 {t_0} + 6 \pi^2 \beta^2{a_1} )}{12 \beta^4}
                      +O(c), ~~
   \cdots,
\eea
with $t_0$, $a_1$ being arbitrary order $c^0$ constants.
On the other hand, if we ask for a more stringent ETH with the condition by requiring $S_A-S_{A,\b}=O(1/c)$, we get
\bea \label{conEENL}
&& \lag T \rag_\r = -\frac{\pi^2 c}{6 \beta^2}+O(1/c), ~~
   \lag \mA \rag_\r = \frac{\pi^4 c^2}{36 \beta^4}+\frac{11 \pi^4 c}{90 \beta^4}+O(c^0), \nn\\
&& \lag \mB \rag_\r = \frac{62 \pi^6 c}{525 \beta^6}+O(c^0), ~~
   \lag \mD \rag_\r = -\frac{\pi^6 c^3}{216 \beta^6}-\frac{11 \pi^6 c^2}{180 \beta^6}+O(c), ~~
   \cdots.
\eea
We call them the next-to-leading order constraints by EE.

Similarly, we can obtain the constraints by requiring the two relative entropies $S(\r_A\|\r_{A,\b})$ or $S(\r_{A,\b}\|\r_{A})$ to be of different orders. If we require  $S(\r_A\|\r_{A,\b}) = O(c^0)$ or $S(\r_{A,\b}\|\r_A) = O(c^0)$, we get exactly the same results as the leading order constraints of EE (\ref{conEEL}).
However, by requiring $S(\r_A\|\r_{A,\b}) = O(1/c)$ or $S(\r_{A,\b}\|\r_A) = O(1/c)$, we get the next-to-leading order constraints of relative entropy
\bea \label{conRENL}
&& \lag T \rag_\r = -\frac{\pi^2 c}{6 \beta^2}
                    +t_0+O(1/c), ~~
   \lag \mA \rag_\r = \frac{\pi^4 c^2}{36 \beta^4}
                     +\frac{(11 \pi^4 - 30\pi^2 \b^2 t_0) c}{90 \beta^4}+O(c^0), \nn\\
&& \lag \mB \rag_\r = \frac{62 \pi^6 c}{525 \beta^6}+O(c^0), ~~
   \lag \mD \rag_\r = -\frac{\pi^6 c^3}{216 \beta^6}
                      -\frac{(11 \pi^6 - 15\pi^4\b^2 t_0 ) c^2}{180 \beta^6}+O(c), ~~
   \cdots,
\eea
{with an arbitrary constant $t_0$ which is order $c^0$.}
They are different from the next-to-leading order constraints of EE (\ref{conEENL}).

As for the expectation values of KdV charges, we require the following quantities to be small,
\be \label{deltaQ}
\f{\lag Q_{2k-1} \rag_\r-\lag Q_{2k-1} \rag_\b}{\lag Q_{2k-1} \rag_\b}, ~~ k=1,2,3,\cdots.
\ee
Note that these quantities do not depend on normalization convention of the KdV charges.
Requiring (\ref{deltaQ}) to be at order $1/c$, we get the leading order constraints of KdV charge expectation values
\bea \label{conKdVL}
&& \lag T \rag_\r = -\frac{\pi^2 c}{6 \beta^2}+O(c^0), ~~
   \lag \mA \rag_\r = \frac{\pi^4 c^2}{36 \beta^4} + O(c), \nn\\
&& \lag \mB \rag_\r = b_2 c^2 + O(c), ~~
   \lag \mD \rag_\r = - \frac{(\pi^6-10 \beta^6 {b_2})c^3 }{216\beta^6}
                      + O(c^2), ~~
   \cdots,
\eea
{with an arbitrary constant $b_2$ which is order $c^0$.}
Requiring (\ref{deltaQ}) to be at order $1/c^2$, we get the next-to-leading order constraints of KdV charge expectation values
\bea \label{conKdVNL}
&& \lag T \rag_\r = - \frac{\pi^2 c}{6 \beta^2} + O(1/c), ~~
   \lag \mA \rag_\r = \frac{\pi^4 c^2}{36 \beta^4}
                    + \frac{11 \pi^4 c}{90 \beta^4}
                    + O(c^0), ~~
   \lag \mB \rag_\r = b_2 c^2 + b_1 c + O(c^0) , \nn\\
&& \lag \mD \rag_\r = - \frac{( \pi^6-10 \beta^6 {b_2})c^3 }{216\beta^6}
                      - \frac{ ( 151\pi^6 - 105 \beta^6 {b_1}-1344 \beta^6 {b_2})c^2}
                             {2268 \beta^6}
                      +O(c), ~~
   \cdots,
\eea
{with arbitrary constants $b_2$ and $b_1$ which are order $c^0$.}

\section{Checks for primary and various descendant eigenstates} \label{secche}

We check the constraints obtained in the previous section for various highly excited energy eigenstates.
The states to be considered include the excited state $|\phi\rag$ with $\phi$ being a primary operator of conformal weight $h_\phi$, and descendant states of the same conformal family $|\td\phi\rag$, $|\p^m\phi\rag$, $|\p^m\td\phi\rag$, where we define the quasiprimary operator $\td \phi := (T\phi) - \f{3}{2(2h_\phi+1)}\p^2\phi$.
Here $m$ is a positive integer.
We also consider the vacuum conformal family descendant states $|\p^m T\rag$, $|\p^m\mA\rag$.
Note that $|\p^m T\rag$ is just a special case of $|\p^m\td\phi\rag$ with $h_\phi=0$.
One can see details of these states in appendix~\ref{appCFT}.

Note that in this section we only check the aforementioned constraints up to  level 6 of the holomorphic sector of the vacuum conformal family. If the constraints are violated, the results are conclusive. If the constraints are satisfied to level 6, we do not know whether the constraints would be violated at higher levels.

We take the thermodynamic limit for the CFT \cite{Lashkari:2016vgj}. This requires that the  conformal weights of the states are of order $L^2$ in $L \to \inf$ and is of order $c$ in large $c$ limit. In the thermodynamic limit, the energy eigenstates we consider fall into three types depending on the values of the parameters.

The type I states include $|\phi\rag$ and $|\td \phi\rag$ with $h_\phi = {H L^2} + o(L^2)$, as well as $|\p^m \phi\rag$ and $|\p^m \td \phi\rag$ with $h_\phi = {H L^2} + o(L^2)$ and $m=o(L^2)$.
{We introduce a positive parameter $H$, which is of $O(L^0)$.}
{Note that, here} $o(L^2)$ denotes the terms satisfying $\lim_{L\to\inf}\f{o(L^2)}{L^2}=0$.
These descendants are {slightly} excited on top of a heavy primary {state}.
Using results (\ref{TABDLphi}) (\ref{TABDLtidlephi}) in appendix~\ref{appCFT}, we get the expectation values
\be
\lag T \rag_{\r_\rI} = -4 \pi^2 H, ~~
\lag \mA \rag_{\r_\rI} = 16 \pi^4 H^2, ~~
\lag \mB \rag_{\r_\rI} = 0, ~~
\lag \mD \rag_{\r_\rI} = -64 \pi^6 H^3, ~~
\cdots,
\ee
where $\rho_\rI$ refers to the type I states. Note that for these states  the contributions from $m$ are suppressed in {$L\to\inf$} limit, the results are ``universal'' , {i.e. it only depends on} the conformal dimension $h_\phi$ of the primary state.
{We see that neither all-order constraint (\ref{conallEERE}) associated with EE/relative entropy nor (\ref{conallKdV}) associated with the expectation values of the KdV charges is} satisfied by the above expectation values for any $H$. However, {the leading constraint (\ref{conEEL}) associated EE and relative entropy is} satisfied with the identification of the parameter
\be
H = \frac{c}{24 \beta^2} + H_0 +O(1/c),
\ee
{with} $t_0=-4\pi^2 H_0$, $a_1=\f{4\pi^4}{3\b^2}H_0$ in (\ref{conEEL}).  Note that $H_0$ is $O(c^0)$ in the $c \to \inf$ limit. Similarly, {the leading order constraint (\ref{conKdVL}) associated with the expectation values of the KdV charges is} satisfied with the identification of the parameter,
\be
H = \frac{c}{24 \beta^2} + O(c^0),
\ee
{with} $b_2=0$ in (\ref{conKdVL}). Furthermore, none of the next-to-leading order {constraints (\ref{conEENL}) for EE , (\ref{conRENL}) for relative entropy  and  (\ref{conKdVNL}) for the expectation values of KdV charges} can be satisfied.

The type II states include $| \p^m \phi \rag$ and $| \p^m \td \phi \rag$ with $h_\phi = {H L^2} + o(L^2)$, $m = {M L^2} + o(L^2)$. {Here we introduce the positive parameters $H$ and $M$, both of which are of $O(L^0)$.}
{These descendants are highly excited on top of a heavy primary state.}
Using the result (\ref{TABDLphi}) and (\ref{TABDLtidlephi}) we get the expectation values of the quasiprimary operators
\bea
&& \lag T \rag_{\r_{\text{II}}} = -4 \pi^2 (H+M), ~~
   \lag \mA \rag_{\r_{\text{II}}} = \frac{8 \pi^4 (H+M) (2 H^2+10 H M+5 M^2)}{H}, ~~
   \lag \mB \rag_{\r_{\text{II}}} = 0, \nn\\
&& \lag \mD \rag_{\r_{\text{II}}} = -\frac{8 \pi^6 (H+M) (8 H^4+112 H^3 M+308 H^2 M^2+252 H M^3+63 M^4)}{H^2}, ~~
   \cdots,
\eea
where ${\r_{\text{II}}}$ refers to the type II states.
These expectation values of quasiprimary operators do not satisfy the all-order constraints (\ref{conallEERE}) or (\ref{conallKdV}). However, they satisfy {the leading order constraint (\ref{conEEL}) for EE and relative entropy}  by the parameters
\be
H = \frac{c}{24 \beta^2} + H_0 +O(1/c), ~~
M = M_0 + O(1/c),
\ee
with $t_0=-4\pi^2 ( H_0 + M_0 )$, $a_1=\f{4\pi^4}{3\b^2} ( H_0 + 3 M_0 )$ in (\ref{conEEL}).
Here $H_0$ and $M_0$ are $O(c^0)$ in the $c \to \inf$ limit.
They also satisfy {the leading order constraint (\ref{conKdVL}) for the expectation values of KdV charges}  by the parameters
\be
H = \frac{c}{24 \beta^2} +O(c^0), ~~
M = O(c^0).
\ee
However, they do not satisfy any of the next-to-leading order constraint (\ref{conEENL}), (\ref{conRENL}), or (\ref{conKdVNL}). Besides, for $M = O(c)$, all the leading order and next-to-leading order constraints are violated.

The type II descendant states with $M = O(c^0)$ are heavily excited on top of a heavy primary {state in the large $L$ expansions}, but are {slightly excited in the large $c$ expansions}. {On the other hand, the} type II descendant states with $M = O(c)$ are heavily excited on top of a heavy primary {state in both the large $L$ and large $c$ expansions. Thus,} we see that the essence for the type II descendant states satisfying the leading order ETH constraints is that the descendant states are {slightly} excited on top of a heavy primary {state in the large $c$ expansions}.

The type III states include states $| \p^m \phi \rag$ and $| \p^m \td \phi \rag$ with $h_\phi=o(L^2)$, $m = {M L^2} + o(L^2)$, as well as $| \p^m T \rag$ and $| \p^m \mA \rag$ with $m = {M L^2} + o(L^2)$.
{Note that $M>0$ and is of $O(L^0)$.}
They are highly excited descendants on top of a light primary {state.}
In thermodynamic limit, such states have divergent expectations values, and would never be close to the canonical ensemble state.

We summarize the results of this section in table~\ref{tab}. The states we have considered can at most match the canonical ensemble state at leading order of large $c$.
For the descendant states  we have considered, to match the thermal state at the leading order of large $c$, it requires that (1) the corresponding primary state matches the canonical ensemble state at the leading order of large $c$; and (2) the conformal weight difference between the primary and descendant states is at most of order $L^2$ in large $L$ limit, and the order $L^2$ part, if non-vanishing, is at most of order $c^0$ in large $c$ limit.
In other words, for the states we have considered, only the descendant states that are {slightly} excited on top of a heavy primary state can satisfy the leading order ETH constraints in the large $c$ limit, and all the ETH constraints are violated for the descendant states that are heavily excited on top of a primary state.

\begin{table}[htbp]
  \centering
\begin{tabular}{|c|c|c|c|c|}
  \hline
  \mlr{2}{type} & \mlr{2}{state}                                                               & \mlc{3}{c|}{constraints to level 6}                                      \\\cline{3-5}
                &                                                                              & leading             & next-to-leading     & all                 \\\hline

  \mlr{4}{I}    & $|\phi\rag$ with $h_\phi = H {L^2} + o(L^2)$                                 & \mlr{4}{\checked}   & \mlr{4}{\texttimes} & \mlr{4}{\texttimes} \\
                & $|\td \phi\rag$ with $h_\phi = H {L^2} + o(L^2)$                             &                     &                     &                     \\
                & $|\p^m\phi\rag$ with $h_\phi = H {L^2} + o(L^2)$, $m = o(L^2)$               &                     &                     &                     \\
                & $|\p^m\td \phi\rag$ with $h_\phi = H {L^2} + o(L^2)$, $m = o(L^2)$           &                     &                     &                     \\\hline

  \mlr{2}{II}   & $|\p^m\phi\rag$ with $h_\phi = H {L^2} + o(L^2)$, $m = M {L^2} + o(L^2)$     & \mlr{2}{\checked}   & \mlr{2}{\texttimes} & \mlr{2}{\texttimes} \\
                & $|\p^m\td \phi\rag$ with $h_\phi = H {L^2} + o(L^2)$, $m = M {L^2} + o(L^2)$ &                     &                     &                     \\\hline

  \mlr{4}{III}  & $|\p^m\phi\rag$ with $h_\phi = o(L^2)$, $m = M {L^2} + o(L^2)$               & \mlr{4}{\texttimes} & \mlr{4}{\texttimes} & \mlr{4}{\texttimes} \\
                & $|\p^m\td \phi\rag$ with $h_\phi = o(L^2)$, $m = M {L^2} + o(L^2)$           &                     &                     &                     \\
                & $|\p^m T\rag$ with $m = M {L^2} + o(L^2)$                                    &                     &                     &                     \\
                & $|\p^m \mA \rag$ with $m = M {L^2} + o(L^2)$                             &                     &                     &                     \\\hline
\end{tabular}
  \caption{The three types primary and descendant states we consider in this section, and whether they satisfy the ETH constraints derived from EE, relative entropy, and the expectation values of KdV charges  up to level 6 of the holomorphic sector of the vacuum conformal family. In the 2nd column we have definitions $H = \frac{c}{24 \beta^2} + O(c^0)$, $M=O(c^0)$, {with the positive constants $H$ and $M$ being of $O(L^0)$.} Note that for $M=O(c)$ all the leading order and next-to-leading order constraints are violated. The 3rd, 4th, and 5th columns all apply to the constraints derived from EE, relative entropy, and the expectation values of KdV charges. We mark \checked\ for constraints that are satisfied and mark \texttimes\ otherwise.}\label{tab}
\end{table}

Due to the limitation of the special descendant states we consider, we do not know whether {our conclusions can apply to the more general} descendant states. {One possibility is that our special descendant states are quite typical, so that our conclusions should be universal, that is, only the descendant states that close enough to their primary state can satisfy ETH at the leading order of large $c$ expansions. On the other hand, if our special descendant states are atypical, then none of conclusion is universal. Thus,
the problems are how typical our special descendant states are, and whether they can represent the majority of the descendant states. These are open questions.}

\section{Conclusion and discussion} \label{seccnd}

We calculated the short interval expansions of EE and relative entropy to order $\ell^{12}$.
Using the results to require ETH of a highly excited state in the context of canonical ensemble state, we got the leading order and next-to-leading order constraints in the large $c$ expansion on the expectation values of the first few quasiprimary operators in the holomorphic sector of the vacuum conformal family.
We also obtained the constraints from the expectation values of the first few KdV charges.

We checked the constraints for the primary and various descendant states. We found that these constraints can only be satisfied for at most the leading order of large $c$, and even the leading order constraints are violated for  highly excited descendant states on top of a primary state.
The essence for the descendant states satisfying the leading order ETH constraints is that the descendant states are slightly excited on top of a heavy primary state in the large $c$ expansions.
Note that when we say a descendant state is close to or far away from its primary state, we refer to just the primary state in the same conformal family of the descendant state.
We stress that we have only considered states with conformal weights $h \pp c L^2$ in the limit of $L\to\inf$ and $c\to\inf$, and so $h \gg c \gg 1$. Both the thermodynamic limit and the large $c$ limit are crucial to our analysis. Since we have only considered some special descendant states, we do not know how the majority of the descendant states might behave.


In \cite{Guo:2018fnv}, we have investigated the conditions for a CFT state to have {{a} bulk classical geometric description in the AdS/CFT correspondence}.  {For a CFT state to be bulk geometric, its leading order EE or R\'enyi entropy calculated in the large $c$ CFT should be the same with the ones calculated by the holographic methods \cite{Ryu:2006bv,Hubeny:2007xt,Dong:2016fnf}. For the heavy states of $O(c)$ conformal weights, the holographic EE and R\'enyi entropy should be at most of $O(c)$ {\cite{Guo:2018fnv}}. By this criterion and the derived conditions, we have found some descendant states to be non-geometric. However,}
There are some differences between the geometric state conditions in \cite{Guo:2018fnv} and the ETH state constraints in this paper as follows.
(1) For the geometric state conditions, the states do not necessarily have translational symmetry and so the one-point functions are not necessarily constants. For the ETH state constraints, we consider the globally excited energy eigenstate, and the one-point functions are constants.
(2) In the geometric state conditions we do not require the state to be at high energy, and so we just focus on the order of large $c$ and do not care about the order of $L$. In the ETH state constraints we need to firstly take thermodynamic limit $L \to \inf$ and then do expansion of large $c$.

Although there are differences mentioned above,  it is still interesting to compare the results. We stress that the comparison in this paragraph is under the thermodynamic limit. For translation-invariant states, the geometric state conditions in \cite{Guo:2018fnv} can be recast as
\bea \label{congeo}
&& \lag T \rag_\r = c t_1 + t_0 + O(1/c), ~~
   \lag \mA \rag_\r = c^2 t_1^2 + c a_0 + O(c^0), \nn\\
&& \lag \mB \rag_\r = O(c), ~~
   \lag \mD \rag_\r = c^3 t_1^3 + 3 c^2 t_1(a_1 - t_1 t_0) + O(c), ~~
   \cdots,
\eea
{with arbitrary constants $t_1$ and $t_0$, which are both of order $c^0$.}
The states satisfying the ETH {constraints (\ref{conEEL}), (\ref{conEENL}), (\ref{conRENL}) associated with EE and relative entropy} also satisfy the geometric state conditions (\ref{congeo}). However, the states satisfying the ETH {constraints (\ref{conKdVL}) and (\ref{conKdVNL}) associated with the  expectation values of KdV charges} do not necessarily satisfy the geometric state conditions.
In fact the geometric state conditions (\ref{congeo}) are not only consistent with but also equivalent to the leading order constraint (\ref{conEEL}) for EE and relative entropy.
Requiring the energy density be non-negative in the thermodynamic limit, we {find that} $t_1$ in (\ref{congeo}) is nonpositive. {Moreover, we can see that the {conditions} (\ref{congeo}) and (\ref{conEEL}) are equivalent if one sets $t_1 = -\frac{\pi^2}{6 \beta^2}$.}
The equivalence of the geometric conditions and the leading order constraints of ETH in the thermodynamic limit is remarkable.
The geometric conditions come from $S_A^{(n)}=O(c)$, which leads to $S_A=O(c)$. The leading order ETH constraints for EE come from $S_A=S_{A,\b}+O(c^0)$ with (\ref{SAb}).
A possible way to understand this is that a translation-invariant geometric state is dual to a Ba\~nados geometry \cite{Banados:1998gg} with a constant stress tensor and in thermodynamic limit the classical Ba\~nados geometry metric is exactly the same as the BTZ black hole metric with some identification of parameters. By the holographic EE \cite{Ryu:2006bv,Hubeny:2007xt}, it ensures that the translation-invariant geometric state satisfies the leading order ETH constraints of EE (\ref{conEEL}).

The ETH state constraints we obtained are about how close a state $\r$ and the canonical ensemble state $\r_\b = \f{1}{Z(\b)}\sum_i \ep^{-\b E_i} |i\rag \lag i|$ with $Z(\b) = \sum_i \ep^{-\b E_i}$ in the thermodynamic limit.
Although the RDM of a primary state is the same as RDM of the thermal state $\r_\b$ in the leading order of $c$, the RDMs of the descendant states we have considered that are far away from the primary state are very different from the thermal state RDM.
Recently we have proved that the RDM of the canonical ensemble state $\r_\b$ and RDM of the microcanonical ensemble state $\r_E = \f{1}{\O(E)}\sum_i \d(E- E_i) |i\rag \lag i|$ with $\O(E) = \sum_i \d(E- E_i)$ are same in thermodynamic limit \cite{Guo:2018djz}.
In both of the canonical and microcanonical ensembles, one needs to average over all the states, including both the primary and descendant states.
In fact, there are far more descendant states than primary states at high energy. The density of all the states is given by the Cardy formula \cite{Cardy:1986ie}
\be
\O(E) \sim \ep^{\sr{\f{2\pi c L E}{3}}},
\ee
and the density of primary states is \cite{Kraus:2016nwo}
\be
\O_p(E) \sim \ep^{\sr{\f{2\pi (c-1) L}{3}\big(E+\f{\pi}{6L}\big)}}.
\ee
Note that in the thermodynamic limit one has $E \to \inf$, $L \to \inf$, and $\ve = E/L$ is finite. We get the ratio of the density of primary states and the density of all states is
\be
\f{\O_p(E)}{\O(E)}  \sim \ep^{ - L \big[ \sr{\f{2\pi c \ve }{3}} - \sr{\f{2\pi (c-1)}{3}\big(\ve+\f{\pi}{6L^2}\big)} \big] }
\to 0 ~\mathrm{as}~ L \to \inf.
\ee
It is exponentially suppressed in the thermodynamic limit. So at a high temperature, or equivalently at a high energy, the average in the canonical or microcanonical ensemble is dominated by descendant states. It is an intriguing puzzle how the average over descendant states behaves like the primary states in large $c$ limit.
Note that we have only considered some very special descendant states in this paper, it is possible most of the descendant states at high levels behave like the primary states at the same  conformal weight.
Another possibility is that each descendant state is different from the primary state at the same conformal weight, while the ensemble average erases the differences.
The techniques used in this paper cannot be applied to a general descendant state.
It would be nice if the issue can be addressed by investigations of general descendant states.

The RDMs of the primary excited state and the canonical ensemble state are same at the leading order of large $c$ \cite{Fitzpatrick:2014vua,Fitzpatrick:2015zha,Asplund:2014coa,Caputa:2014eta}, but they are in fact different with subleading corrections of $1/c$ \cite{He:2017vyf,Basu:2017kzo,He:2017txy,Lashkari:2017hwq}.
In this paper we obtained the same results for some descendant states that are close to their primary states.
A possible resolution to the mismatch of the excited states and the thermal state was proposed and investigated in \cite{He:2017vyf,Basu:2017kzo,He:2017txy,Lashkari:2017hwq}, and it is to replace the canonical ensemble with the GGE \cite{Rigol:2006}. In 2D CFT there are infinite number of conserved charges that commute with the Hamiltonian \cite{Sasaki:1987mm,Eguchi:1989hs,Bazhanov:1994ft}, which are just the KdV charges.
A strong form of ETH in the context of GGE is that the RDM a typical energy eigenstate is same as the RDM of GGE state in the thermodynamic limit. A weak form is to require the state to be an eigenstate of all KdV charges.
The chemical potentials of the GGE state are determined by requiring that the GGE state has the same KdV charges as the excited state.
As can be seen in (\ref{SrAsA}) and figure~\ref{gencon}, two general states that have the same KdV charges do not necessarily have the same RDMs. It is an open question whether ETH in the context of GGE is correct for the 2D large $c$ CFT.

\section*{Acknowledgments}

We would like to thank Pasquale Calabrese, Chong-Sun Chu, Hong Liu, Erik Tonni for helpful discussions.
We thank the anonymous referee for valuable comments and suggestions.
WZG would like to thank the physics department of Jinan University for hospitality where part of this work was completed.
JZ would like to thank the Galileo Galilei Institute for Theoretical Physics and the organisers of the workshop ``Entanglement in Quantum Systems'' for hospitality and to thank participants of the workshop for helpful discussions.
WZG is supported in part by the National Center of Theoretical Science (NCTS).
FLL is supported by Taiwan Ministry of Science and Technology through Grant No.~103-2112-M-003-001-MY3.
JZ is supported by Fondazione Cariplo and Regione Lombardia through Grant No.~2015-1253,
and by ERC Consolidator Grant No.~771536 (NEMO).

\appendix

\section{Quasiprimary operators and their correlation functions} \label{appCFT}

In this appendix, we give some useful details of 2D CFT we need in this paper, including both reviews and some new calculations. The basics of 2D CFT can be found in \cite{Ginsparg:1988ui,DiFrancesco:1997nk,Blumenhagen:2009zz}.

\subsection{Vacuum conformal family}

Many details of the quasiprimary operators in the vacuum conformal family can be found in the papers \cite{Chen:2013dxa,Li:2016pwu,He:2017vyf}. We only consider the holomorphic sector, and the anti-holomorphic sector is similar. We count the number of independent holomorphic operators at each level in the vacuum conformal family as
\be \label{trxL0}
\tr x^{L_0} = \prod_{k=2}^{+\inf} \f{1}{1-x^k},
\ee
among which the holomorphic quasiprimary operators are counted as
\be \label{x1xtrxL0}
x + (1-x) \tr x^{L_0} = 1+x^2+x^4+2 x^6+3 x^8+x^9+4 x^{10}+2 x^{11}+7 x^{12}+3 x^{13}+O(x^{14}).
\ee
We list these quasiprimary operators in table~\ref{tab1}, and to level 9 their explicit forms can be found in \cite{Chen:2013dxa,Li:2016pwu,He:2017vyf}.

\begin{table}[htbp]\centering\begin{tabular}{|c|c|c|c|c|c|c|c|c|c|c|c|}\hline
level        & 0 & 2   & 4     & 6            & 8                   & 9           & 10             & 11             & 12             & 13             & $\cdots$ \\\hline
quasiprimary & 1 & $T$ & $\mA$ & $\mB$, $\mD$ & $\mE$, $\mH$, $\mI$ & $\mA^{(9)}$ & $\mA^{(10,m)}$ & $\mA^{(11,m)}$ & $\mA^{(12,m)}$ & $\mA^{(13,m)}$ & $\cdots$ \\
\hline\end{tabular}
\caption{Holomorphic quasiprimary operators in the vacuum conformal family. The ranges in which the $m$'s take values can be read in (\ref{x1xtrxL0}). For examples, at level 10 we have $m=1,2,3,4$, and at level 11 we have $m=1,2$.}\label{tab1}
\end{table}

At level 0, it is trivially the identity operator $1$.
At level 2, it is the stress tensor $T$ with the usual normalization $\a_T=\f{c}{2}$.
At level 4 we have the quasiprimary operator and its normalization factor
\be \label{Adef}
\mA=(TT)-\f{3}{10}\p^2T, ~~ \a_{\mc A}=\f{c(5c+22)}{10}.
\ee
At level 6, we have the orthogonalized quasiprimary operators
\bea \label{BDdef}
&& \mB=(\p T\p T)-\f{4}{5}(\p^2TT)-\f{1}{42}\p^4T, \nn\\
&& \mD= (T(TT))-\f{9}{10}(\p^2TT)-\f{1}{28}\p^4 T + \f{93}{70c+29} \mc B,
\eea
and their normalization factors are
\be
\a_{\mc B}=\frac{36c (70 c+29)}{175}, ~~
\a_{\mc D}=\frac{3 c (2 c-1) (5 c+22) (7 c+68)}{4 (70 c+29)}.
\ee
We do not need the explicit forms of other quasiprimary operators in this paper. The conformal transformation rules of the quasiprimary operators to level 8 can be found in \cite{He:2017vyf}.

In the vacuum conformal family, one can define an infinite number of mutually commuting conserved KdV charges $Q_{2k-1}$, $k=1,2,3,\cdots$ \cite{Sasaki:1987mm,Eguchi:1989hs,Bazhanov:1994ft}. In terms of the quasiprimary operators, we can write the first three KdV charges on a cylinder as \cite{Guo:2018fnv}
\bea \label{KdVgeneral}
&& Q_1 = \f{1}{(2\pi\ii)^2} \int_0^L \f{d w}{L} T(w) , ~~
   Q_3 = \f{1}{(2\pi\ii)^4} \int_0^L \f{d w}{L} \Big[ \mA(w) + \f{3}{10} \p^2 T(w) \Big], \nn\\
&& Q_5 = \f{1}{(2\pi\ii)^6} \int_0^L \f{d w}{L} \Big[ \mD
         -\frac{25 (2 c+7) (7 c+68)}{108 (70 c+29)} \mB
         - \frac{2 c - 23}{108} \p^2\mA
         - \frac{c-14}{280} \p^4 T \Big].
\eea

For the CFT on a circle with period $L$ in the ground state, which is just a vertical cylinder with spatial period $L$, we have the one point functions \cite{Chen:2013kpa,Chen:2013dxa}
\bea \label{TABDL}
&& \lag T \rag_L=\f{\pi^2c}{6L^2}, ~~
   \lag \mc A \rag_L=\f{\pi^4c(5c+22)}{180L^4}, \\
&& \lag \mc B \rag_L=-\frac{62 \pi^6 c}{525 L^6}, ~~
   \lag \mc D \rag_L=\f{\pi^6 c(2c-1)(5c+22)(7c+68)}{216(70c+29)L^6}. \nn 
\eea
For the CFT on a circle with period $L$ in thermal state with inverse temperature $\b \gg L$, which is a torus with module $\t=\ii\b/L$, we use the results in \cite{Chen:2016lbu} and get the one point functions expanded by $q=\ep^{-2\pi\b/L} \ll 1$
\bea\label{TABDLq}
&& \lag T \rag_{L,q} = \frac{\pi^2}{6 L^2} [ c-48 q^2-72 q^3-144 q^4+O(q^5) ], \nn\\
&& \lag \mA \rag_{L,q} = \frac{\pi^4}{180 L^4} [ c (5 c+22)+480 (5 c+22) q^2+2160 (5 c+22) q^3+30240 (c+6) q^4+O(q^5) ], \nn\\
&& \lag \mB \rag_{L,q} = -\frac{2 \pi^6}{525 L^6} [ 31 c-1008 (120 c+1) q^2-1512 (720 c+161) q^3-3024 (1640 c+841) q^4+O(q^5) ], \nn\\
&& \lag \mD \rag_{L,q} = \frac{\pi^6 (2 c-1) (7 c+68)}{216 (70 c+29) L^6}
                         [ c (5 c+22)+1584 (5 c+22) q^2+6696 (5 c+22) q^3 \nn\\
&& \phantom{\lag \mD \rag_{L,q} =}
                          +432 (215 c-638) q^4+O(q^5) ].
\eea
Note that, as stated in \cite{Chen:2016lbu}, the above one point functions can be expanded to an arbitrary order of $q$.

For the CFT on an infinite straight line in canonical ensemble state with inverse temperature $\b$, which is just a horizontal cylinder with temporal period $\b$, we have \cite{Chen:2013kpa,Chen:2013dxa}
\bea \label{TABDb}
&& \lag T \rag_\b=-\f{\pi^2c}{6\b^2}, ~~
   \lag \mc A \rag_\b=\f{\pi^4c(5c+22)}{180\b^4}, \nn\\
&& \lag \mc B \rag_\b=\frac{62 \pi^6 c}{525 \b^6}, ~~
   \lag \mc D \rag_\b=-\f{\pi^6 c(2c-1)(5c+22)(7c+68)}{216(70c+29)\b^6}.
\eea
Then we use (\ref{KdVgeneral}), and get the expectation values of KdV charges for the thermal state
\be \label{KdVb}
\lag Q_1 \rag_\b = \frac{c}{24 \beta^2} , ~~
\lag Q_3 \rag_\b = \frac{c (5 c+22)}{2880 \beta^4} , ~~
\lag Q_5 \rag_\b = \frac{c (3 c+14) (7 c+68)}{290304 \beta^6}. 
\ee

We have orthogonalized the operators, such that the correlation function of two holomorphic quasiprimary operators $\mX_{1,2}$ on a complex plane $\rC$ with coordinate $f$ takes the form
\be \label{twopt}
\lag {\mX_1}(f_1){\mX_2}(f_2) \rag_\rC = \f{\d_{{\mX_1}{\mX_2}}\a_{\mX_1}}{f_{12}^{2h_{\mX_1}}},
\ee
where we have defined $f_{12} \equiv f_1-f_2$. The correlation function of three holomorphic quasiprimary operators $\mX_{1,2,3}$ takes the form
\be \label{threept}
\lag \mX_1(f_1)\mX_2(f_2)\mX_3(f_3) \rag_\rC = \f{C_{\mX_1\mX_2\mX_3}}
                                           {f_{12}^{h_{\mX_1}+h_{\mX_2}-h_{\mX_3}}
                                            f_{13}^{h_{\mX_1}+h_{\mX_3}-h_{\mX_2}}
                                            f_{23}^{h_{\mX_2}+h_{\mX_3}-h_{\mX_1}}},
\ee
with $C_{\mX_1\mX_2\mX_3}$ being the structure constant. In this paper, we need the structure constants
\bea \label{CXYZ}
&& C_{TTT}=c, ~~
   C_{TT\mA}=\frac{c (5 c+22)}{10}, ~~
   C_{TT\mB}=-\frac{2 c (70 c+29)}{35}, \nn\\
&& C_{TT\mD}=0, ~~
   C_{T\mA\mA}=\frac{2 c (5 c+22)}{5}, ~~
   C_{T\mA\mB}=-\frac{24 c (5 c+22)}{25}, \nn\\
&& C_{T\mA\mD}=\frac{3 c (2 c-1) (5 c+22) (7 c+68)}{4 (70 c+29)}, ~~
   C_{\mA\mA\mA}=\frac{c (5 c+22) (5 c+64)}{25}, \nn\\
&& C_{\mA\mA\mB}=-\frac{4 c (5 c+22) (14 c+73)}{35}, ~~
   C_{\mA\mA\mD}=\frac{6 c (2 c-1) (5 c+22) (7 c+68)}{70 c+29}.
\eea
These structure constants can be calculated by using Virasoro algebra. For examples we have $C_{TT\cA}=\lag0|L_2L_2\cA_{-4}|0\rag$, $C_{T\cA\mB}=\lag0|L_2\cA_{4}\mB_{-6}|0\rag$. Note that an operator $\cO(f)$ with conformal weight $h$ on a complex plane can be expanded in terms of its modes $\cO_m$ as $\mO(f)=\sum_m \f{\cO_m}{f^{h+m}}$. As usual, the modes of $T$ are denoted by $L_m$ and satisfy the Virasoro algebra. The modes of $\mA$, $\mB$, $\mD$ can be written in terms of $L_m$.

We also need the four-point functions
\bea \label{mptfn1}
&& \lag T(f_1)T(f_2)T(f_3)T(f_4) \rag_\rC = \f{c^2}{4}\Big( \f{1}{f_{12}^4f_{34}^4} + \cdots \Big)_3
                                          + c \Big( \f{1}{f_{12}^2f_{23}^2f_{34}^2f_{41}^2} + \cdots \Big)_3, \nn\\
&& \lag T(f_1)T(f_2)T(f_3)\mA(f_4) \rag_\rC = \f{c(5c+22)}{20}
   \Big[ \Big( \f{f_{12}^2}{f_{13}^2f_{23}^2f_{14}^4f_{24}^4} + \cdots \Big)_3
     + 2 \Big( \f{1}{f_{12}^2f_{14}^2f_{24}^2f_{34}^4}+ \cdots \Big)_3
   \Big], \nn\\
&& \lag T(f_1)T(f_2)T(f_3)\mB(f_4) \rag_\rC = -\f{2c(70c+29)}{35} \Big( \f{f_{12}^4}{f_{13}^2f_{23}^2f_{14}^6f_{24}^6} + \cdots \Big)_3
                                              + \f{36c(4c-1)}{5} \f{1}{f_{14}^4f_{24}^4f_{34}^4}, \nn\\
&& \lag T(f_1)T(f_2)T(f_3)\mD(f_4) \rag_\rC = \frac{3 c (2 c-1) (5 c+22) (7 c+68)}{4 (70 c+29)} \f{1}{f_{14}^4f_{24}^4f_{34}^4}, \nn\\
&& \lag T(f_1)T(f_2)\mA(f_3)\mA(f_4) \rag_\rC = \f{c^2(5c+22)}{20}\Big[ \f{1}{f_{12}^4f_{34}^8}
                                                                      + 2 \Big( \f{1}{f_{13}^4f_{24}^4f_{34}^4} + \cdots \Big)_2
                                                                  \Big] \nn \\
&& \phantom{\lag T(f_1)T(f_2)\mA(f_3)\mA(f_4) \rag_\rC =}
   + \f{c(5c+22)}{25} \Big[ \f{9f_{12}^4}{f_{13}^4f_{14}^4f_{23}^4f_{24}^4} - \f{28}{f_{13}^2f_{14}^2f_{23}^2f_{24}^2f_{34}^4} \\
&& \phantom{\lag T(f_1)T(f_2)\mA(f_3)\mA(f_4) \rag_\rC =}
                          - 8 \Big( \f{f_{12}^2}{f_{13}^2f_{24}^2f_{34}^2f_{14}^4f_{23}^4} + \cdots \Big)_2
                          + 10 \Big( \f{f_{13}^2f_{24}^2}{f_{12}^2f_{14}^4f_{23}^4f_{34}^6} + \cdots \Big)_2
                      \Big], \nn
\eea
the five-point functions
\bea \label{mptfn2}
&& \lag T(f_1)T(f_2)T(f_3)T(f_4)T(f_5) \rag_\rC = \f{c^2}{2}\Big( \f{1}{f_{12}^2f_{23}^2f_{31}^2f_{45}^4} + \cdots \Big)_{10}
                                              + c \Big( \f{1}{f_{12}^2f_{23}^2f_{34}^2f_{45}^2f_{51}^2} + \cdots \Big)_{12}, \nn\\
&& \lag T(f_1)T(f_2)T(f_3)T(f_4)\mA(f_5) \rag_\rC = \f{c(5c+22)}{20} \Big[
      c \Big( \f{1}{f_{12}^4f_{35}^4f_{45}^4} + \cdots \Big)_6
    + 4 \Big( \f{1}{f_{12}^2f_{23}^2f_{15}^2f_{35}^2f_{45}^4} + \cdots \Big)_{12} \nn\\
&& ~~ ~~ ~~ ~~ ~~ ~~ ~~ ~~ 
    + 4 \Big( \f{1}{f_{12}^2f_{34}^2f_{15}^2f_{25}^2f_{35}^2f_{45}^2} + \cdots \Big)_{3}
    + 2 \Big( \f{f_{12}^2f_{34}^2}{f_{13}^2f_{14}^2f_{23}^2f_{24}^2f_{15}^2f_{25}^2f_{35}^2f_{45}^2} + \cdots \Big)_{3}
                                                                     \Big],
\eea
and the six-point function
\bea \label{mptfn3}
&& \lag T(f_1)T(f_2)T(f_3)T(f_4)T(f_5)T(f_6) \rag_\rC =
       \f{c^3}{8} \Big( \f{1}{f_{12}^4f_{34}^4f_{56}^4} + \cdots \Big)_{15}
     + c^2 \Big( \f{1}{f_{12}^2f_{23}^2f_{31}^2f_{45}^2f_{56}^2f_{64}^2} + \cdots \Big)_{10} \nn\\
&& ~~ ~~ ~~ ~~ ~~ ~~ ~~ ~~
     + \f{c^2}{2} \Big( \f{1}{f_{12}^2f_{23}^2f_{34}^2f_{41}^2f_{56}^4} + \cdots \Big)_{45}
     + c \Big( \f{1}{f_{12}^2f_{23}^2f_{34}^2f_{45}^2f_{56}^2f_{61}^2} + \cdots \Big)_{60}.
\eea
In the above multi-point functions, we have used the notation in the form
\be ( * + \cdots )_\#, \ee
where $\cdots$ denotes permutation terms of $*$ and $\#$ is the total number of terms in the parentheses. For example, the five-point function $\lag T(f_1)T(f_2)T(f_3)T(f_4)\mA(f_5) \rag_\rC$ is invariant under the permutations of $f_1,f_2,f_3,f_4$, and for it we have
\be
\Big( \f{1}{f_{12}^4f_{35}^4f_{45}^4} + \cdots \Big)_6 \equiv \f{1}{f_{12}^4f_{35}^4f_{45}^4}
                                                            + \f{1}{f_{13}^4f_{25}^4f_{45}^4}
                                                            + \f{1}{f_{14}^4f_{25}^4f_{35}^4}
                                                            + \f{1}{f_{23}^4f_{15}^4f_{45}^4}
                                                            + \f{1}{f_{24}^4f_{15}^4f_{35}^4}
                                                            + \f{1}{f_{34}^4f_{15}^4f_{25}^4}.
\ee

The above multi-point functions (\ref{mptfn1}), (\ref{mptfn2}), (\ref{mptfn3}) are derived from the two-point and three-point functions (\ref{twopt}), (\ref{threept}), the conformal Ward identity, and the OPEs
\bea
&& T(f)T(0) = \f{c}{2f^4} + \f{2T(0)}{f^2} + \f{\p T(0)}{f} + \cdots, \nn\\
&& T(f)\mA(0) = \f{(5c+22)T(0)}{5f^4} + \f{4\mA(0)}{f^2} + \f{\p \mA(0)}{f} + \cdots, \nn\\
&& T(f)\mB(0) = - \frac{4(70 c+29)}{35f^6} \Big( T(0) - \f{f\p T(0)}{2} + \f{f^2\p^2T(0)}{20} \Big)
                - \f{48\mA(0)}{5f^4} + \f{6\mB(0)}{f^2} + \f{\p \mB(0)}{f} + \cdots, \nn\\
&& T(f)\mD(0) = \frac{15 (2 c-1) (7 c+68)\mA(0)}{2 (70 c+29)f^4} + \f{6\mD(0)}{f^2} + \f{\p \mD(0)}{f} + \cdots.
\eea
The form the OPE of two quasiprimary operators is fixed by the conformal symmetry, and all we need to do is to fill in the correct structure constants that we already know (\ref{CXYZ}).

\subsection{Non-vacuum conformal family}

We consider the non-vacuum conformal family of the primary operator $\phi$, with conformal weight $h_\phi>0$ and normalization $\a_\phi$. Without loss of generality and for simplicity, we choose it to be holomorphic, and the generalization to a non-chiral primary operator is straightforward. The next quasiprimary operator in the conformal family of $\phi$ is
\be
\td\phi = (T\phi) - \f{3}{2(2h_\phi+1)}\p^2\phi,
\ee
with conformal weight $h_{\td\phi} = h_\phi+2$ and the normalization factor
\be
\a_{\td\phi} = \frac{\alpha_{\phi} [ 16 h_{\phi }^2 + 2 (c-5) h_{\phi }+c ]}
                    {2 (2 h_{\phi}+1)}.
\ee

Using the state operator correspondence in 2D CFT {and radial quantization on the complex plane},
we construct the ket and bra states {on the cylinder}
\bea
&& |\phi\rag = \phi(0)|0\rag, ~~
   \lag \phi | = \lag0|\phi(\inf) = \lim_{z \to 0} \lag 0 | z^{-2 h_\phi} \phi(z^{-1}), \nn\\
&& |\p^m\phi\rag = \p^m\phi(0)|0\rag, ~~
   \lag \p^m \phi | = \lag0|\p^m\phi(\inf) = \lim_{z \to 0} \lag 0 | \p_z^m [ z^{-2 h_\phi} \phi(z^{-1}) ].
\eea
Similarly we define $|\td\phi\rag$, $\lag \td\phi |$, $|\p^m\td\phi\rag$, $\lag \p^m \td\phi |$.
The state $|\phi\rag$ is primary, and states $|\td\phi \rag$, $|\p^m\phi\rag$, $|\p^m\td\phi\rag$ are descendants.
We have the normalization factors
\bea
&& \a_{\p^m\phi} = \lag \p^m \phi | \p^m \phi \rag = \f{m!(2h_\phi+m-1)!}{(2h_\phi-1)!}\a_\phi,  \nn\\
&& \a_{\p^m\td\phi} = \lag \p^m \td \phi | \p^m \td \phi \rag = \f{m!(2h_\phi+m+3)!}{(2h_\phi+3)!}\a_{\td\phi}.
\eea
Using these states, we construct the normalized density matrices of the whole system
\bea \label{states}
&& \r_{L,\phi} = \f{|\phi\rag\lag\phi|}{\a_\phi}, ~~
   \r_{L,\p^m\phi} = \f{|\p^m\phi\rag\lag\p^m\phi|}{\a_{\p^m\phi}}, \nn\\
&& \r_{L,\td\phi} = \f{|\td\phi\rag\lag\td\phi|}{\a_{\td\phi}}, ~~
   \r_{L,\p^m\td\phi} = \f{|\p^m\td\phi\rag\lag\p^m\td\phi|}{\a_{\p^m\td\phi}}.
\eea

As in \cite{He:2017vyf}, we use the Virasoro algebra and get the structure constants
\bea
&& C_{\phi\phi T}= \a_\phi h_\phi, ~~
   C_{\phi\phi\mA}=\f{ \a_\phi h_\phi(5h_\phi+1)}{5}, ~~
   C_{\phi\phi\mB}=-\frac{2 \a_\phi  h_{\phi}(14 h_{\phi}+1)}{35} ,  \nn\\
&& C_{\phi\phi\mD}=\frac{\a_\phi h_{\phi} [(70 c+29) h_{\phi}^2+(42 c-57) h_{\phi}+2(4 c-1)]}{70 c+29} .
\eea
From the structure constants and the conformal transformation rules of $T$, $\mA$, $\mB$, $\mD$, we get the expectation values
\bea \label{TABDLphi}
&& \lag T \rag_\phi = - \frac{\pi^2 ( 24 h_\phi -c )}{6 L^2}, ~~
   \lag\mA \rag_\phi = \frac{\pi^4  [ 2880 h_\phi^2 -240 (c+2) h_\phi + c (5 c+22) ]}{180 L^4}, \nn\\
&& \lag\mB \rag_\phi = \frac{2 \pi^6  ( 504 h_\phi - 31 c )}{525 L^6}, \nn\\
&& \lag\mD \rag_\phi = - \frac{\pi^6}{216 (70 c+29) L^6} \big[
                                      13824 (70 c+29) h_{\phi}^3
                                    - 1728 (c+4) (70 c+29) h_{\phi}^2 \nn\\
&& \phantom{ \lag \mD \rag_\phi =}
                                    + 72 (70 c^3+617c^2+938c-248) h_{\phi}
                                    - c (2 c-1) (5 c+22) (7 c+68) \big].
\eea
Similarly, we get the structure constants
\bea
&& C_{\td\phi\td\phi T} = \a_{\td \phi}(h_{\phi }+2 ), ~~
   C_{\td\phi\td\phi \mA} = \frac{\a_{\td \phi} [ 10 h_{\phi }^3+127 h_{\phi }^2+5(2 c+3) h_{\phi }+ 5c+22 ]}{5 (2 h_{\phi }+1)}, \nn\\
&& C_{\td\phi\td\phi \mB} = -\frac{2 \a_{\td \phi} [28 h_{\phi }^3
                                  +2368 h_{\phi }^2
                                  +(280 c-1227) h_{\phi }
                                  +2(70c+29)]}{35  (2 h_{\phi }+1 )}, \nn\\
&& C_{\td\phi\td\phi \mD} = \frac{\a_{\td \phi}h_{\phi }}{(70 c+29)  (2 h_{\phi }+1 )}
                                           [ 2(70c+29) h_{\phi }^3
                                            +(4354 c+1655) h_{\phi }^2 \nn\\
&& \phantom{C_{\td\phi\td\phi \mD} =}
                                            +(420 c^2+568 c-12421 )h_{\phi }
                                            + 210 c^2+1103c+7558 ].
\eea
Then we get the expectations values
\bea\label{TABDLtidlephi}
&& \lag T \rag_{\td \phi} = - \frac{\pi^2 ( 24 h_\phi - c + 48 )}{6 L^2}, \nn\\
&& \lag \mA \rag_{\td \phi} = \frac{\pi ^4 [ 5760 h_{\phi }^3-480 (c-148) h_{\phi }^2+2 (5 c^2+2302 c+1680) h_{\phi }+(c+480) (5c+22) ]}
                                   {180(2 h_{\phi }+1) L^4}, \nn\\
&& \lag \mB \rag_{\td \phi} = \frac{2 \pi ^6 [ 1936368 h_ {\phi }^2 + 2 (120929 c - 603540) h_ {\phi } + 120929 c + 1008 ]}
                                   {525(2 h_{\phi }+1) L^6}, \nn
\eea\bea
&& \lag \mD \rag_{\td \phi} = - \frac{\pi ^6}{216 (70 c+29)(2 h_{\phi }+1) L^6}
                                [ 27648 (70 c+29) h_{\phi }^4
                                 -3456 (c-240) (70 c+29) h_{\phi }^3 \nn\\
&& \phantom{\lag \mD \rag_{\td \phi} =}
                                 +144 ( 70c^3+19937c^2-139570c-1090280) h_{\phi }^2
                                 -2 ( 70 c^4+109313c^3-324562c^2\nn\\
&& \phantom{\lag \mD \rag_{\td \phi} =}-6990752c-52852896)h_{\phi }
                                 -(c+1584) (2 c-1) (5 c+22) (7 c+68)  ].
\eea

For general quasiprimary operators $\mX$, $\mY$,  from (\ref{twopt}) and (\ref{threept}) we obtain the three-point function on a complex plane
\be
\f{\lag \p^m\mY(\inf)\p^p\mX(f)\p^m\mY(0) \rag_\rC}{\lag \p^m\mY(\inf)\p^m\mY(0) \rag_\rC}
= \f{C_{\mY\mX\mY}}{\a_\mY}\f{1}{f^{h_\mX+p}} \f{(-)^pp!C_{h_\mX+p-1}^p}{C_{2h_\mY+m-1}^m}
  \sum_{i=0}^m (C_{h_\mX+i-1}^i)^2 C_{2h_\mY-h_\mX+m-i-1}^{m-i},
\ee
with the binomial coefficient $C_x^y = \G(x+1)/[\G(y+1)\G(x-y+1)]$. Using the three-point function we get the expectations values $\lag \mX \rag_{\p^m\phi}$, $\lag \mX \rag_{\p^m\td\phi}$ with $\mX=T,\mA,\mB,\mD$. Similarly, we obtain the expectations values $\lag \mX \rag_{\p^m T}$, $\lag \mX \rag_{\p^m \mA}$ with $\mX=T,\mA,\mB,\mD$. We will not give the explicit forms of these expectations values.

\section{Short interval expansions of EE and relative entropy}\label{appexp}

We review the short interval expansions of EE and relative entropy by OPE of twist operators. We also obtain the EE and relative entropy to higher orders than the ones in literature. The method of twist operators was proposed in \cite{Calabrese:2004eu} to calculate R\'enyi entropy in 2D CFT,  and was latter refined in \cite{Cardy:2007mb}. The OPE of twist operators was formulated in \cite{Headrick:2010zt,Calabrese:2010he,Chen:2013kpa}. The replica trick of calculating the relative entropy in CFT was developed in \cite{Lashkari:2014yva,Lashkari:2015dia}.
As shown in \cite{Lashkari:2017hwq}, for the subsystem ETH of highly excited states in a 2D chaotic CFT, we only need to consider the contributions from the vacuum conformal family in the OPE of twist operators. {This is based on the observation that evaluation of EE or R\'enyi entropy for the large $c$ CFTs are dominated by the saddle of vacuum conformal family so that the results are universal \cite{Hartman:2013mia}.}
Without loss of generality, we only include the contributions from the holomorphic sector, and the contributions from the anti-holomorphic sector can be added easily.
Note that the results of EE (\ref{EEgeneral}) and relative entropy (\ref{relativegeneral}) can be  applied to general translation-invariant states, because in the derivations we use nothing but the fact that the one-point functions are constants. We also stress that our results of EE (\ref{EEgeneral}) and relative entropy (\ref{relativegeneral}) serve well for the investigation of subsystem ETH in the 2D large $c$ CFT, but for the translation-invariant states in a concrete CFT, e.g. the Ising model, we need to add the contributions from other operators in the non-vacuum conformal families.

\subsection{OPE of twist operators}

We consider one short interval $A=[0,\ell]$ on a general Riemann surface $\mR$ with translational symmetry, and the constant time slice is in a state with density matrix $\r$.
Tracing the degrees of freedom of the complement of $A$ that we call $\bar A$, we get the RDM $\r_A=\tr_{\bar A}\r$. To get the EE
\be
S_A = - \tr_A ( \r_A \log\r_A ),
\ee
we use the replica trick and first calculate the R\'enyi entropy
\be
S_A^{(n)} = -\f{\log\tr_A\r_A^n}{n-1},
\ee
and then take the $n \to 1$ limit. We calculate the partition function $\tr_A\r_A^n$ of the CFT on an $n$-fold Riemann surface $\mR^n$, and it equals the two-point function of twist operators $\s$, $\td\s$ in the $n$-fold CFT on a single copy of the Riemann surface $\mR$ \cite{Calabrese:2004eu}
\be \label{trArAn}
\tr_A\r_A^n = \lag \s(\ell)\td\s(0) \rag_\mR.
\ee
Note that in this paper we just focus on the holomorphic sector of the twist operators, and the addition of the anti-holomorphic sector is straightforward.

The twist operators are primary operators with conformal weights \cite{Calabrese:2004eu}
\be
h_\s=h_{\td \s}=\f{c(n^2-1)}{24n}.
\ee
We can write the OPE of twist operators as \cite{Headrick:2010zt,Calabrese:2010he,Chen:2013kpa}
\be \label{sztdsw}
\s(z)\td \s(w)=\f{c_n}{(z-w)^{2h_\s}} \sum_K d_K \sum_{p=0}^\inf \f{c_K^p}{p!}(z-w)^{h_K+p} \p^p\Phi_K(w),
\ee
with $c_n$ being the normalization factor, the summation $K$ being over all the orthogonalized holomorphic quasiprimary operators $\Phi_K$ in $\CFT^n$, and $h_K$ being the conformal weight of $\Phi_K$. We have defined
\be
c_K^p \equiv \f{C_{h_K+p-1}^p}{C_{2h_K+p-1}^p},
\ee
with $C_x^y$ denoting the binomial coefficient. The OPE coefficient $d_K$ can be calculated as \cite{Calabrese:2010he}
\be \label{dKdef}
d_K = \f{1}{\a_K\ell^{h_K}}\lim_{z \to \inf} z^{2h_K}\lag \Phi_K(z) \rag_{\mR_{n,1}},
\ee
with $\a_K$ being the normalization of $\Phi_K$. We have used $\mR_{n,1}$ to denote the $n$-fold Riemann surface $\mR^n$ that results from the replica trick for one interval $A=[0,\ell]$ on the complex plane $\rC$. To calculate the right-hand side of (\ref{dKdef}), we map $\mR_{n,1}$ with coordinate $z$ to a complex plane with coordinate $f$ by the transformation \cite{Calabrese:2004eu,Calabrese:2010he}
\be \label{fz}
f(z) = \Big(\f{z-\ell}{z}\Big)^{1/n}.
\ee

Because of the translational symmetry, all the one-point functions on $\mR$ are constants. Then we use (\ref{sztdsw}) and write (\ref{trArAn}) as
\be
\tr_A\r_A^n = \f{c_n}{\ell^{2h_\s}} \sum_K d_K\ell^{h_K}\lag\Phi_K\rag_\mR,
\ee
with the summation of $K$ being over all the $\CFT^n$ holomorphic quasiprimary operators $\Phi_K$ that are direct products of the holomorphic quasiprimary operators in each copy of the original CFT.
In this paper we only consider the contributions from the holomorphic sector of the vacuum conformal family, and the relevant $\CFT^n$ quasiprimary operators are counted as
\bea \label{x1xtrxL0n}
&& [x+(1-x)\tr x^{L_0}]^n =
1
+n x^2
+\frac{n(n+1)}{2} x^4
+\frac{n(n^2+3 n+8)}{6}x^6
+\frac{n (n+1) (n^2+5 n+30)}{24} x^8  \nn\\
&& \phantom{[x+(1-x)\tr x^{L_0}]^n =}
+n x^9
+\frac{n (n+1) (n+2) (n^2+7 n+72)}{120} x^{10}
+n(n+1) x^{11}  \nn\\
&& \phantom{[x+(1-x)\tr x^{L_0}]^n =}
+\frac{n (n+3)(n^4+12 n^3+169 n^2+438 n+640)}{720} x^{12} \nn\\
&& \phantom{[x+(1-x)\tr x^{L_0}]^n =}
+\f{n(n+1)(n+2)}{2} x^{13}
+O(x^{14}).
\eea
Note the definition of $\tr x^{L_0}$ in (\ref{trxL0}).
The direct product quasiprimary operators take the form
\be
\Phi_K^{j_1\cdots j_k} = \mX_1^{j_1}\cdots \mX_k^{j_k},
\ee
and we list these operators {up} to conformal weight 13 in table~\ref{tab2}.

\begin{table}[htbp]
  \centering
\begin{tabular}{|c|c|c|c|c|c|c|c|c|c|c|}\cline{1-5}\cline{7-11}
  CW       & quasiprimary           & ?                   & \#              & \#
   && CW         & quasiprimary                 & ?                   & \#                      & \#                                                            \\ \cline{1-5} \cline{7-11}
  \mlr{1}{0}  & 1                      & -                   & 1               & 1                                                 && \mlr{2}{11}  & $\mA^{(11,m)}$               & \texttimes          & $2n$                    & \mlr{2}{$n(n+1)$}                                             \\ \cline{1-5} \cline{8-10}
  \mlr{1}{2}  & $T$                    & \checked            & $n$             & $n$                                               &&              & $T\mA^{(9)}$                 & \texttimes          & $n_2$                   &                                                               \\ \cline{1-5} \cline{7-11}
  \mlr{2}{4}  & $\mA$                  & \texttimes          & $n$             & \mlr{2}{$\f{n(n+1)}{2}$}                          && \mlr{13}{12} & $\mA^{(12,m)}$               & \texttimes          & $7n$                    & \rot{13}{$\frac{n (n+3)(n^4+12 n^3+169 n^2+438 n+640)}{720}$} \\ \cline{2-4} \cline{8-10}
              & $TT$                   & \checked            & $\f{n_2}{2}$    &                                                   &&              & $T\mA^{(10,m)}$              &                     &                         &                                                               \\ \cline{1-5}
  \mlr{3}{6}  & $\mB$, $\mD$           & \texttimes          & $2n$            & \rot{3}{$\frac{n(n^2+3 n+8)}{6}$}                 &&              & $\mA\mE$, $\mA\mH$, $\mA\mI$ & \texttimes          & $8n_2$                  &                                                               \\ \cline{2-4}
              & $T\mA$                 & \texttimes          & $n_2$           &                                                   &&              & $\mB\mD$                     &                     &                         &                                                               \\ \cline{2-4} \cline{8-10}
              & $TTT$                  & \checked            & $\f{n_3}{6}$    &                                                   &&              & $\mB\mB$, $\mD\mD$           & \checked            & $n_2$                   &                                                               \\ \cline{1-5} \cline{8-10}
  \mlr{5}{8}  & $\mE$, $\mH$, $\mI$    & \texttimes          & $3n$            & \rot{5}{$\frac{n (n+1) (n^2+5 n+30)}{24}$}        &&              & $TT\mE$, $TT\mH$             & \mlr{2}{\texttimes} & \mlr{2}{$\f{3n_3}{2}$}  &                                                               \\ \cline{2-4}
              & $T\mB$, $T\mD$         & \texttimes          & $2n_2$          &                                                   &&              & $TT\mI$                      &                     &                         &                                                               \\ \cline{2-4} \cline{8-10}
              & $\mA\mA$               & \checked            & $\f{n_2}{2}$    &                                                   &&              & $T\mA\mB$, $T\mA\mD$         & \mlr{2}{\checked}   & \mlr{2}{$\f{13n_3}{6}$} &                                                               \\ \cline{2-4}
              & $TT\mA$                & \checked            & $\f{n_3}{2}$    &                                                   &&              & $\mA\mA\mA$                  &                     &                         &                                                               \\ \cline{2-4} \cline{8-10}
              & $TTTT$                 & \checked            & $\f{n_4}{24}$   &                                                   &&              & $TTT\mB$, $TTT\mD$           & \mlr{2}{\checked}   & \mlr{2}{$\f{7n_4}{12}$} &                                                               \\ \cline{1-5}
  \mlr{1}{9}  & $\mA^{(9)}$            & \texttimes          & $n$             & $n$                                               &&              & $TT\mA\mA$                   &                     &                         &                                                               \\ \cline{1-5} \cline{8-10}
  \mlr{7}{10} & $\mA^{(10,m)}$         & \texttimes          & $4n$            & \rot{7}{$\frac{n (n+1) (n+2) (n^2+7 n+72)}{120}$} &&              & $TTTT\mA$                    & \checked            & $\f{n_5}{24}$           &                                                               \\ \cline{2-4} \cline{8-10}
              & $T\mE$, $T\mH$, $T\mI$ & \mlr{2}{\texttimes} & \mlr{2}{$5n_2$} &                                                   &&              & $TTTTTT$                     & \checked            & $\f{n_6}{720}$          &                                                               \\ \cline{7-11}
              & $\mA\mB$, $\mA\mD$     &                     &                 &                                                   && \mlr{4}{13}  & $\mA^{(13,m)}$               & \texttimes          & $3n$                    & \rot{4}{$\f{n(n+1)(n+2)}{2}$}                                 \\ \cline{2-4} \cline{8-10}
              & $TT\mB$, $TT\mD$       & \texttimes          & $n_3$           &                                                   &&              & $T\mA^{(11,m)}$              & \mlr{2}{\texttimes} & \mlr{2}{$3n_2$}         &                                                               \\ \cline{2-4}
              & $T\mA\mA$              & \checked            & $\f{n_3}{2}$    &                                                   &&              & $\mA\mA^{(9)}$               &                     &                         &                                                               \\ \cline{2-4} \cline{8-10}
              & $TTT\mA$               & \checked            & $\f{n_4}{6}$    &                                                   &&              & $TT\mA^{(9)}$                & \texttimes          & $\f{n_3}{2}$            &                                                               \\ \cline{2-4} \cline{7-11}
              & $TTTTT$                & \checked            & $\f{n_5}{120}$  &                                                   && \mlc{5}{c|}{$\cdots$}                                                                                                                                       \\ \cline{1-5} \cline{7-11}
\end{tabular}
\caption{The $\CFT^n$ holomorphic quasiprimary operators that are direct products of the vacuum conformal family holomorphic quasiprimary operators in each copy of the original CFT.
We use CW to denote the conformal weight.
We have omitted the replica indices for these operators.
For example, at conformal weight 8, $TT\mA$ denotes $T_{j_1}T_{j_2}\mA_{j_3}$ with $0\leq j_{1,2,3} \leq n-1$, $j_1 < j_2$, $j_1 \neq j_3$, $j_2 \neq j_3$.
In the third column, we mark \checked\ for nonidentity operators with generally non-vanishing contributions to the single interval EE in a translation-invariant state, i.e. with non-vanishing coefficients $a_K$ defined in (\ref{aX1dddX2def}), and we mark \texttimes\ for operators with vanishing coefficients $a_{\mX_1\cdots\mX_k}$. Note that for $k=0$, i.e. the identity operator, we do not need to calculate the coefficient $a_{\mX_1\cdots\mX_k}$.
We count the number of operators in the fourth and fifth columns, with the notation $n_k=n(n-1)\cdots(n-k+1)$. The counting  is consistent with (\ref{x1xtrxL0n}).}
\label{tab2}
\end{table}

The one-point function of $\mX_1^{j_1}\cdots \mX_k^{j_k}$ on $\mR$ is independent of the replica indices
\be
\lag \mX_1^{j_1}\cdots \mX_k^{j_k} \rag_\mR = \lag \mX_1 \rag_\mR \cdots \lag \mX_k \rag_\mR.
\ee
We sum the replica indices of the OPE coefficient $d_{\mX_1\cdots\mX_k}^{j_1\cdots j_k}$ and define \cite{Chen:2016lbu}
\be \label{bX1dddXk}
b_{\mX_1\cdots\mX_k} \equiv \sum_{j_1,\cdots,j_k} d_{\mX_1\cdots\mX_k}^{j_1\cdots j_k}
\textrm{~with~some~constraints~for~}0 \leq j_1,\cdots,j_k \leq n-1.
\ee
The constraints are to avoid overcounting of the quasiprimary operators. For example, the constraints for $d_{TT\mA}^{j_1j_2j_3}$ are $j_1 < j_2$, $j_1 \neq j_3$, $j_2 \neq j_3$. Except the identity operator, all the coefficients $b_{\mX_1\cdots\mX_k}$ are vanishing in the limit $n \to 1$. We get the R\'enyi entropy
\be
S_A^{(n)} = \f{c(n+1)}{12n} \log\f{\ell}{\e} - \f{1}{n-1}\log\Big( 1 + \sum_{k=1}^n \sum_{\{\mX_1, \cdots, \mX_k\}} \ell^{h_{\mX_1} + \cdots + h_{\mX_k}} b_{\mX_1\cdots\mX_k} \lag\mX_1\rag_\mR \cdots \lag\mX_k\rag_\mR \Big).
\ee
From the coefficient $b_{\mX_1\cdots\mX_k}$ with $k \geq 1$, we further define \cite{Chen:2017ahf,He:2017txy}
\be \label{aX1dddX2def}
a_{\mX_1\cdots\mX_k} \equiv - \lim_{n \to 1} \f{b_{\mX_1\cdots\mX_k}}{n-1},
\ee
and get the EE written as
\be
S_A = \f{c}{6} \log\f{\ell}{\e} + \sum_{k=1}^\inf \sum_{\{\mX_1, \cdots, \mX_k\}} \ell^{h_{\mX_1} + \cdots + h_{\mX_k}} a_{\mX_1\cdots\mX_k} \lag\mX_1\rag_\mR \cdots \lag\mX_k\rag_\mR.
\ee

In this paper we focus on the entanglement entropy, instead of the R\'enyi entropy. To calculate the coefficients $a_{\mX_1\cdots\mX_k}$ we do not need the full forms of $b_{\mX_1\cdots\mX_k}$ or $d_{\mX_1\cdots\mX_k}^{j_1\cdots j_k}$.
As can be seen in (\ref{aX1dddX2def}), to calculate $a_{\mX_1\cdots\mX_k}$ we can omit the $O(n-1)^2$ part of $b_{\mX_1\cdots\mX_k}$.
A general holomorphic quasiprimary operator $\mX$ transforms under a general map $z \to f(z)$ as
\be \label{Xz}
\mX(z) = f'^{h_\mX} \mX(f) + \cdots,
\ee
with $\cdots$ denoting terms with the Schwarzian derivative
\be \label{Sch}
s(z) = \f{f'''(z)}{f'(z)} -\f32 \Big(\f{f''(z)}{f'(z)}\Big)^2.
\ee
For the transformation (\ref{fz}) we have
\be \label{sz}
s(z) = \f{(n^2-1)\ell^2}{2n^2z^2(z-\ell)^2}.
\ee
When we calculate $d_{\mX_1\cdots\mX_k}^{j_1\cdots j_k}$ using (\ref{dKdef}), the contributions from $\cdots$ terms in (\ref{Xz}) would be of order $O(n-1)$ or of higher orders.
For $k \geq 2$, when we compute $b_{\mX_1\cdots\mX_k}$ using (\ref{dKdef}), the summation of the replica indices would lead to another order $O(n-1)$ or higher order factor for each term.
For $k \geq 2$, the $\cdots$ terms in (\ref{Xz}) only contribute order $O(n-1)^2$ or higher order terms to $b_{\mX_1\cdots\mX_k}$, and so would not contribute to $a_{\mX_1\cdots\mX_k}$ defined in (\ref{aX1dddX2def}).
For $k=1$, $a_\mX=0$ for $h_\mX>2$. We will prove it the next appendix.

Up to conformal weight 13 the nonidentity $\CFT^n$ quasiprimary operators with non-vanishing $a_{\mX_1\cdots\mX_k}$ are marked with \checked\ in the 3rd column of table~\ref{tab2}.
For $k=1$ we only need to calculate $a_T$.
For $k \geq 2$, we use the various multi-point functions in appendix~\ref{appCFT} and get $d_{\mX_1\cdots\mX_k}^{j_1\cdots j_k}$ up to the omitted irrelevant $O(n-1)$ terms. Summing the replica indices, we get $b_{\mX_1\cdots\mX_k}$ up to some omitted irrelevant $O(n-1)^2$ terms. Then we get $a_{\mX_1\cdots\mX_k}$. The non-vanishing coefficients $a_{\mX_1\cdots\mX_k}$ are listed as follows
\bea\label{aK}
&& a_T = -\f16, ~~
   a_{TT} = -\frac{1}{30 c}, ~~
   a_{TTT} = -\frac{4}{315 c^2}, \nn\\
&& a_{\mA\mA} = -\frac{1}{126 c (5 c+22)}, ~~
   a_{TT\mA} = \frac{1}{315 c^2}, ~~
   a_{TTTT} = -\frac{c+8}{630 c^3}, \nn\\
&& a_{T\mA\mA} = -\frac{16}{693 c^2 (5 c+22)}, ~~
   a_{TTT\mA} = \frac{32}{3465 c^3}, ~~
   a_{TTTTT} = -\frac{16 (c+5)}{3465 c^4}, \nn\\
&& a_{\mB\mB} = -\frac{25}{123552 c (70 c+29)}, ~~
   a_{\mD\mD} = -\frac{70 c+29}{18018 c (2 c-1) (5 c+22) (7 c+68)}, \nn\\
&& a_{T\mA\mB} = -\frac{10}{1287 c^2 (70 c+29)}, ~~
   a_{T\mA\mD} = \frac{5}{3003 c^2 (5 c+22)}, ~~
   a_{\mA\mA\mA} = \frac{4 (5 c+64)}{3003 c^2 (5 c+22)^2}, \nn\\
&& a_{TTT\mB} = \frac{5 (14 c+43)}{9009 c^3 (70 c+29)}, ~~
   a_{TTT\mD} = -\frac{2}{9009 c^3}, ~~
   a_{TT\mA\mA} = -\frac{585 c+10804}{90090 c^3 (5 c+22)},  \nn\\
&& a_{TTTT\mA} = \frac{2 (33 c+784)}{45045 c^4}, ~~
   a_{TTTTTT} = -\frac{2 (11 c^2+380c+1480)}{45045 c^5}.
\eea
Up to conformal weight 8, the coefficients $a_{\mX_1\cdots\mX_k}$ have been calculated in \cite{He:2017txy} using the results in \cite{Chen:2013dxa,Chen:2016lbu}, and at  conformal weight 10 and conformal weight 12 the results here are new. We have used the coefficients up to conformal weight 10 to calculate the Holevo information in \cite{Guo:2018djz}.

\subsection{EE}

Using the above coefficients, we get EE of a short interval $A=[0,\ell]$ in a general translation-invariant state $\r$ on a Riemann surface $\mR$ \cite{He:2017txy}
\bea \label{EEgeneral}
&& S_A = \f{c}{6}\log\f\ell\e + \ell^2 a_T \Trho
                                  + \ell^4 a_{TT} \Trho^2
                                  + \ell^6 a_{TTT} \Trho^3 \nn\\
&& \phantom{S_A =}
           + \ell^8 \big( a_{\mA\mA} \Arho^2
                        + a_{TT\mA} \Trho^2\Arho
                        + a_{TTTT} \Trho^4 \big) \nn\\
&& \phantom{S_A =}
           + \ell^{10} \big( a_{T\mA\mA} \Trho\Arho^2
                           + a_{TTT\mA} \Trho^3\Arho
                           + a_{TTTTT} \Trho^5 \big) \nn\\
&& \phantom{S_A =}
           + \ell^{12} \big( a_{\mB\mB} \Brho^2
                           + a_{\mD\mD} \Drho^2
                           + a_{T\mA\mB} \Trho\Arho\Brho
                           + a_{T\mA\mD} \Trho\Arho\Drho \nn\\
&& \phantom{S_A =}
                           + a_{\mA\mA\mA} \Arho^3
                           + a_{TTT\mB} \Trho^3\Brho
                           + a_{TTT\mD} \Trho^3\Drho
                           + a_{TT\mA\mA} \Trho^2\Arho^2 \nn\\
&& \phantom{S_A =}
                           + a_{TTTT\mA} \Trho^4\Arho
                           + a_{TTTTTT} \Trho^6  \big) + O(\ell^{14}).
\eea

To check the EE formula and the coefficients $a_{\mX_1 \cdots\mX_k}$ in (\ref{aK}), we consider several examples.  The first case is that $\mR$ is a vertical cylinder with spatial period $L$. We denote the state density matrix as $\r_L$, we have the expectation values (\ref{TABDL}).
We get the entanglement entropy
\be
S_{A,L} = \f{c}{6}\log\f\ell\e -\frac{\pi^2 c \ell^2}{36 L^2}-\frac{\pi^4 c \ell^4}{1080 L^4}-\frac{\pi^6 c \ell^6}{17010
   L^6}-\frac{\pi^8 c \ell^8}{226800 L^8}-\frac{\pi^{10} c \ell^{10}}{2806650 L^{10}}-\frac{691 \pi^{12}
   c \ell^{12}}{22986463500 L^{12}}+O(\ell^{14}),
\ee
which matches the exact result \cite{Calabrese:2004eu}
\be
S_{A,L} = \f{c}{6} \log \Big( \f{L}{\pi\e} \sin\f{\pi\ell}{L} \Big).
\ee

It is similar for the CFT on an infinite straight line in thermal state with inverse temperature $\b$, which is just the 2D CFT on a horizontal cylinder with temporal period $\b$. We use the expectation values (\ref{TABDb}) and get a result that matches the EE (\ref{SAb}).

On a torus with low temperature, we have spatial and temporal periods $L$ and $\b$ that satisfy $L\ll \b$. We denote the density matrix as $\r_{L,q}$ with $q=\ep^{-2\pi\b/L} \ll 1$. On the low temperature torus we have the one-point functions (\ref{TABDLq}), putting which in (\ref{EEgeneral}) we get the EE
\bea
&& S_{A,L,q} = \f{c}{6} \log\f{\ell}{\e}+ \Big[ -\frac{c}{36}+\frac{4 q^2}{3}+2 q^3+4 q^4+O(q^5) \Big] \Big( \f{\pi\ell}{L} \Big)^2
                + \Big[ -\frac{c}{1080}
                        +\frac{4 q^2}{45}
                        +\frac{2q^3}{15} \nn\\
&& \phantom{S_{A,L,q} =}
                        +\frac{4 (c-8) q^4}{15 c}
                        +O(q^5) \Big] \Big( \f{\pi\ell}{L} \Big)^4
                + \Big[ -\frac{c}{17010}
                        +\frac{8 q^2}{945}
                        +\frac{4 q^3}{315}
                        +\frac{8 (c-16) q^4}{315 c} \nn\\
&& \phantom{S_{A,L,q} =}
                        +O(q^5) \Big] \Big( \f{\pi\ell}{L} \Big)^6
                + \Big[ -\frac{c}{226800}
                        +\frac{4 q^2}{4725}
                        +\frac{2 q^3}{1575}
                        -\frac{4 (159 c+728) q^4}{1575 c}
                        +O(q^5) \Big] \Big( \f{\pi\ell}{L} \Big)^8 \nn\\
&& \phantom{S_{A,L,q} =}
                + \Big[ -\frac{c}{2806650}
                        +\frac{8 q^2}{93555}
                        +\frac{4 q^3}{31185}
                        -\frac{104 (295 c+1312) q^4}{155925 c}
                        +O(q^5) \Big] \Big( \f{\pi\ell}{L} \Big)^{10} \nn\\
&& \phantom{S_{A,L,q} =}
                + \Big[ -\frac{691 c}{22986463500}
                        +\frac{5528 q^2}{638512875}
                        +\frac{2764 q^3}{212837625}
                        -\frac{8 (21728429 c+15283768) q^4}{212837625 c}  \nn\\
&& \phantom{S_{A,L,q} =}
                        +O(q^5) \Big] \Big( \f{\pi\ell}{L} \Big)^{12} +O(\ell^{14}),
\eea
and it is consistent with the exact result
\bea \label{SrALq}
&& S_{A,L,q} = \f{c}{6} \log \Big( \f{L}{\pi\e}\sin\f{\pi\ell}{L} \Big)
              + \Big( 1 - \f{\pi\ell}{L} \cot \f{\pi\ell}{L} \Big) 2 ( 2q^2 + 3q^3 + 6q^4 )  \nn\\
&& \phantom{ S_{A,L,q} =} + \Big[
   -\f{64}{315} \sin^8 \f{\pi\ell}{L}
   -\f{1}{192} \f{1}{\cos^6\f{\pi\ell}{L}} \Big( 76
                                               + 87 \cos \f{2\pi\ell}{L}
                                               + 44 \cos \f{4\pi\ell}{L}
                                               + 3 \cos \f{6\pi\ell}{L}
                                            \Big) \nn\\
&& \phantom{ S_{A,L,q} =}
   +\f{1}{32} \f{\pi\ell}{L} \cot \f{\pi\ell}{L} \f{1}{\cos^8\f{\pi\ell}{L}} \Big( 9
                                                                                 + 18 \cos \f{2\pi\ell}{L}
                                                                                 + 6 \cos \f{4\pi\ell}{L}
                                                                                 + 2 \cos \f{6\pi\ell}{L}
                                                                              \Big) \nn\\
&& \phantom{ S_{A,L,q} =}
   -\f{1}{15c} \f{1}{\cos^2\f{\pi\ell}{L}} \Big( 97
                                               + 59 \cos \f{2\pi\ell}{L}
                                               - 7 \cos \f{4\pi\ell}{L}
                                               + \cos \f{6\pi\ell}{L}
                                           \Big) \nn\\
&& \phantom{ S_{A,L,q} =}
   +\f{2}{c} \f{\pi\ell}{L} \cot \f{\pi\ell}{L} \f{1}{\cos^4\f{\pi\ell}{L}} \Big( 2
                                                                                + 3 \cos \f{2\pi\ell}{L}
                                                                             \Big)
   \Big] q^4 + O(q^5),
\eea
which is valid as long as the interval length $\ell$ is not comparable with total length $L$.
The order $c^0$ part of (\ref{SrALq}) was calculated to order $q^2$ in \cite{Cardy:2014jwa}, to order $q^3$ in \cite{Chen:2014unl}, and to order $q^4$ in \cite{Chen:2015uia}. The order $1/c$ part of (\ref{SrALq}) is new, and we calculate it using the method in \cite{Cardy:2014jwa,Chen:2014unl,Chen:2015uia}.

For the CFT on a cylinder with spatial period $L$ in the primary state $|\phi\rag$, we denote the density matrix as $\r_{L,\phi}$. We have the expectation values (\ref{TABDLphi}), from which we get the EE
\bea \label{SALphi}
&& S_{A,L,\phi} = \f{c}{6}\log\f\ell\e
               +\frac{\pi ^2\ell ^2( 24 h_{\phi} - c) }{36 L^2}
               -\frac{\pi ^4\ell ^4( 24 h_{\phi} - c)^2 }{1080 c L^4}
               +\frac{\pi ^6\ell ^6( 24 h_{\phi} -c)^3 }{17010 c^2L^6 } \nn\\
&& \phantom{S_{A,L,\phi} =}
  -\frac{\pi ^8 \ell ^8}
        {226800 c^3 (5 c+22) L^8}
\big[ 184320 (9 c+88) h_{\phi }^4
 -92160 (3 c+22) c h_{\phi }^3  \nn\\
&& \phantom{S_{A,L,\phi} =}
 +1152 (15 c+82) c^2 h_{\phi }^2
 -96 (5 c+22) c^3 h_{\phi }
 +(5 c+22) c^4 \big] \nn\\
&& \phantom{S_{A,L,\phi} =}
  +\frac{\pi ^{10}\ell ^{10} ( 24h_{\phi }-c )}
        {2806650 c^4 (5 c+22) L^{10}}
\big[ 552960 (3 c+110)h_{\phi }^4
 -276480 (c+22) c h_{\phi }^3  \nn\\
&& \phantom{S_{A,L,\phi} =}
 +3456 (5 c+54) c^2 h_{\phi }^2
 -96 (5 c+22) c^3 h_{\phi }
 +(5 c+22) c^4 \big] \nn\\
&& \phantom{S_{A,L,\phi} =}
   -\frac{\pi ^{12}\ell ^{12}}
         {22986463500 c^5 (2 c-1) (5 c+22)^2(7 c+68) L^{12}}   \\
&& \phantom{S_{A,L,\phi} =}
\times \big[ 16721510400 ( 2764 c^4+430763c^3+6713346c^2+20890232c-12177440) h_{\phi }^6 \nn\\
&& \phantom{S_{A,L,\phi} =}
 -1393459200 ( 8292 c^4+917833c^3+13434350c^2+40315616c-23630816) c h_{\phi}^5 \nn\\
&& \phantom{S_{A,L,\phi} =}
 +5806080 ( 207300c^4+15298019c^3+204391942c^2+582309160c-344044096) c^2 h_{\phi }^4 \nn\\
&& \phantom{S_{A,L,\phi} =}
 -276480 ( 241850 c^4+11090729c^3+127175130c^2+332835448c-199092032) c^3 h_{\phi }^3 \nn\\
&& \phantom{S_{A,L,\phi} =}
 +1728 (5 c+22) ( 241850 c^3+5525383c^2+32112238c-17278696) c^4 h_{\phi}^2 \nn\\
&& \phantom{S_{A,L,\phi} =}
 -99504 (2 c-1) (5 c+22)^2 (7 c+68) c^5 h_{\phi }
 +691 (2 c-1) (5 c+22)^2 (7c+68) c^6 \big]
   +O(\ell^{14}). \nn
\eea
The result to order $\ell^8$ has been calculated in \cite{He:2017vyf}. Setting $h_\phi = \f{c}{24} \big( \f{L^2}{\b^2} + 1 \big)$ in (\ref{SALphi}), we get a result that matches (\ref{SAb}) are order $O(c)$ in large $c$ limit, and this is consistent with the exact result in \cite{Asplund:2014coa,Caputa:2014eta}.

\subsection{Relative entropy}

It is similar to the relative entropy.
For two general translation-invariant states $\r$ and $\s$, one can define the relative entropy of the RDMs $\r_A$ and $\s_A$ as follows:
\be
S(\r_A\|\s_A) = \tr_A( \r_A\log\r_A ) - \tr_A( \r_A\log\s_A ).
\ee
The replica trick of calculating the relative entropy in CFT was developed in \cite{Lashkari:2014yva,Lashkari:2015dia}, and one takes the $n \to 1$ limit of the quantity
\be
S_n(\r_A\|\s_A) = \f{1}{n-1}\log\f{\tr_A\r_A^n}{\tr_A(\r_A\s_A^{n-1})}.
\ee
Using the OPE of twist operators described in the previous subsections, we get the relative entropy \cite{He:2017txy}
\bea \label{relativegeneral}
&& \hspace{-6mm}
   S(\r_A\|\s_A) =
   -\ell^4 a_{{TT}} (\Trho-\Tsig)^2
   -\ell^6 a_{{TTT}} (\Trho-\Tsig)^2 (\Trho+2 \Tsig) \nn\\
&& \hspace{2mm}
   -\ell^8 \big[ a_{\mc{A}\mc{A}} (\Arho-\Asig)^2
               +a_{{TT\mc{A}}} (\Trho-\Tsig)
                               (\Trho\Arho+\Tsig\Arho - 2 \Tsig \Asig) \nn\\
&& \hspace{2mm}
               +a_{{TTTT}} (\Trho-\Tsig)^2 (\Trho^2 + 2 \Trho\Tsig + 3\Tsig^2) \big] \nn\\
&& \hspace{2mm}
   -\ell^{10} \big[  a_{{T\mc{A}\mc{A}}} (\Trho\Arho+\Trho\Asig-2 \Tsig\Asig)(\Arho-\Asig) \nn\\
&& \hspace{2mm}
                    +a_{{TTT\mc{A}}}(\Trho-\Tsig)( \Trho^2\Arho + \Trho\Tsig\Arho + \Tsig^2\Arho - 3\Tsig^2\Asig)\nn\\
&& \hspace{2mm}
                    +a_{{TTTTT}} (\Trho-\Tsig)^2 ( \Trho^3 + 2 \Trho^2\Tsig + 3\Trho\Tsig^2 + 4\Tsig^3) \big] \nn\\
&& \hspace{2mm}
-\ell^{12} \big[
   a_{{\mB\mB}} (\Brho-\Bsig)^2 + a_{{\mD\mD}} (\Drho-\Dsig)^2 \nn\\
&& \hspace{2mm}
  -a_{T\mA\mB}(\Trho\Asig\Bsig+\Tsig\Arho\Bsig+\Tsig\Asig\Brho-\Trho\Arho\Brho-2\Tsig\Asig\Bsig)  \nn\\
&& \hspace{2mm}
  -a_{T\mA\mD}(\Trho\Asig\Dsig+\Tsig\Arho\Dsig+\Tsig\Asig\Drho-\Trho\Arho\Drho-2\Tsig\Asig\Dsig) \nn\\
&& \hspace{2mm}
  +a_{{\mA\mA\mA}} (\Arho-\Asig)^2 (\Arho+2 \Asig) \nn\\
&& \hspace{2mm}
  +a_{TTT\mB}(\Trho-\Tsig) (\Trho^2\Brho + \Trho\Tsig\Brho + \Tsig^2\Brho - 3\Tsig^2\Bsig) \nn\\
&& \hspace{2mm}
  +a_{TTT\mD}(\Trho-\Tsig) (\Trho^2\Drho + \Trho\Tsig\Drho + \Tsig^2\Drho - 3\Tsig^2\Dsig) \nn\\
&& \hspace{2mm}
  -a_{TT\mA\mA}( 2\Trho\Tsig\Asig^2+2\Tsig^2\Arho\Asig-\Trho^2\Arho^2-3\Tsig^2\Asig^2) \\
&& \hspace{2mm}
  +a_{TTTT\mA}(\Trho-\Tsig)( \Trho^3\Arho+\Trho^2\Tsig\Arho+\Trho\Tsig^2\Arho+\Tsig^3\Arho-4\Tsig^3\Asig) \nn\\
&& \hspace{2mm}
  +a_{TTTTTT}(\Trho-\Tsig)^2(\Trho^4+2\Trho^3\Tsig+3\Trho^2\Tsig^2+4\Trho\Tsig^3+5\Tsig^4) \nn
\big] +O(\ell^{14}).
\eea

For $\r_{A,L_1}$ and $\r_{A,L_2}$ we use (\ref{TABDL}) and get
\bea
&& S(\r_{A,L_1}\|\r_{A,L_2}) = \frac{\pi^4 c (L_1^2-L_2^2)^2 \ell^4}{1080 L_1^4 L_2^4}
                            +\frac{\pi^6 c (2 L_1^6-3 L_2^2 L_1^4+L_2^6) \ell^6}{17010 L_1^6 L_2^6} \nn\\
&& \phantom{S(\r_{A,L_1}\|\r_{A,L_2}) =}
                            +\frac{\pi^8 c (3 L_1^8-4 L_2^2 L_1^6+L_2^8) \ell^8}{226800 L_1^8 L_2^8}
                            +\frac{\pi^{10} c (4 L_1^{10}-5 L_2^2 L_1^8+L_2^{10}) \ell^{10}}{2806650 L_1^{10} L_2^{10}} \nn\\
&& \phantom{S(\r_{A,L_1}\|\r_{A,L_2}) =}
                            +\frac{691 \pi^{12} c (5 L_1^{12}-6 L_2^2 L_1^{10}+L_2^{12}) \ell^{12}}{22986463500 L_1^{12} L_2^{12}}
                            +O(\ell^{14}),
\eea
which is consistent with the exact result \cite{Sarosi:2016oks,Sarosi:2016atx}
\be
S(\r_{A,L_1}\|\r_{A,L_2}) = \frac{c}{6} \log \frac{L_2 \sin\frac{\pi  \ell}{L_2}}{L_1 \sin\frac{\pi  \ell}{L_1}}
                         +\frac{c}{12} \Big(1-\frac{L_2^2}{L_1^2}\Big) \Big( 1-\frac{\pi\ell}{L_2}\cot\frac{\pi\ell}{L_2}\Big).
\ee

\section{Proof of $a_\mX=0$ for $h_\mX>2$} \label{apppro}

In this appendix, we give a proof of $a_\mX=0$ for $\mX$ being a quasiprimary operator in the holomorphic sector of the vacuum conformal family and $h_\mX>2$. Note that $h_\mX>2$ is equivalent to $h_\mX\geq4$. General $a_{\mX_1\cdots\mX_k}$ is defined in (\ref{aX1dddX2def}), and $a_\mX$ is just the special $k=1$ case.

Under a general conformal transformation $z \to f(z)$, the operator $\mX$ transforms formally as
\be
\mX(z) = \sum_\mY \sum_p F_{\mX,\p^p\mY}[f(z)]\p^p\mY(f(z)),
\ee
with the coefficients $F_{\mX,\p^p\mY}[f(z)]$ being composed by derivatives of $f(z)$ and the summation of $\mY$ being over all holomorphic quasiprimary operators including the identity operator 1.
For example, for the stress tensor $T$, it is
\be
T(z) = f'(z)^2 T(f(z)) + \f{c}{12}s(z),
\ee
with the Schwarzian derivative (\ref{Sch}).
We focus on the coefficient with $\mY=1$, and define
\be
F_{\mX}(z) = F_{\mX,1}[f(z)].
\ee
For examples
\be
F_T(z) = \f{c}{12}s(z), ~~ F_\mA(z) = \f{c(5c+22)}{720}s(z)^2.
\ee
We have
\be
a_\mX = - \f{1}{\a_\mX \ell^{h_\mX}} \lim_{n\to1, z \to\inf} \f{z^{2h_\mX}F_\mX(z)}{n-1},
\ee
with the conformal transformation (\ref{fz}). Note that (\ref{sz}), $s(z) = O(n-1)$. To prove $a_\mX = 0$ for $h_\mX>2$, we only need to show $F_\mX = O(s^2)$ for a small $s$.

All the operators in holomorphic sector of the vacuum conformal family can be constructed from $T$ by derivatives, normal orderings, and linear combinations. We can recursively organize all general holomorphic quasiprimary operators $\{\mX\}$ as linear combinations of operators in the forms $(\p^p T \mX)$, $\p^q \mX$ with integers $p=0,1,2,\cdots$, $q=1,2,3,\cdots$.
Note the relation $(\p^p T \p^r \mX) = \p (\p^p T \p^{r-1} \mX) - (\p^{p+1} T \p^{r-1} \mX)$ for $r \geq 1$, we do not need the include $(\p^p T \p^r \mX)$ with $r\geq1$.
For examples, at level 2 we have $T$, at level 4 we have $(TT)$, $\p^2 T$ and get $\mA$, and at level 6 we have $(\p^2 TT)$, $(T\mA)$, $\p^4 T$, $\p^2 \mA$ and get $\mB$, $\mD$. Explicitly, we can recursively write a quasiprimary operator $\mX$ with $h_\mX>2$ as
\be \label{dec}
\mX = \sum_\mY [ u_{\mX\mY} (\p^{h_\mX-h_\mY-2}T\mY) + v_{\mX\mY} \p^{h_\mX-h_\mY}\mY ],
\ee
where the summation is over all the nonidentity quasiprimary operators $\mY$ with $h_\mY \leq h_\mX-1$ and $u_{\mX\mY}$, $v_{\mX\mY}$ are constants. In fact the constants $u_{\mX\mY}=0$, for $h_\mY\geq h_{\mX}-1$, $v_{\mX\mY}=0$ for $h_\mY\geq h_\mX$. Generally for a fixed $\mX$, the decomposition (\ref{dec}) may not be unique. Writing in terms of states, we have
\be
|\mX\rag = \sum_\mY [ u_{\mX\mY} (h_\mX-h_\mY-2)! L_{-h_\mX+h_\mY} + v_{\mX\mY} L_{-1}^{h_\mX-h_\mY} ] |\mY\rag
\ee
with $L_k$ being modes of the stress tensor $T$.
We multiply it with the bra state
\be
\lag \p^{h_\mX-2} T | = (h_\mX-2)! \lag0|L_{h_\mX}.
\ee
Using the orthogonality of the quasiprimary operators and the Virasoro algebra we get
\be
v_{\mX T} = - \f{12u_{\mX T}}{(h_\mX-3)(h_\mX-2)h_\mX(h_\mX+1)}.
\ee
We write $\mX$ as
\bea \label{in}
&& \mX = \sum_{\mY\neq T} [ u_{\mX\mY} (\p^{h_\mX-h_\mY-2}T\mY) + v_{\mX\mY} \p^{h_\mX-h_\mY}\mY ] \\
&& \phantom{\mX =}
         + u_{\mX T} \Big[ (\p^{h_\mX-4}TT) - \f{12}{(h_\mX-3)(h_\mX-2)h_\mX(h_\mX+1)} \p^{h_\mX-2} T \Big]. \nn
\eea
Note that for the holomorphic nonidentity quasiprimary operator $\mY\neq T$, we have $h_\mY\geq4$.

The normal ordering operator can be written as
\be
(\p^p T\mY)(w) = \f{1}{2\pi\ii}\oint_w \f{d z}{z-w} \p^p T(z) \mY(w).
\ee
Note that $\mY$ is a quasiprimary operator, and at least $F_\mY=O(s)$. From the conformal transformations of $T$ and $\mY$ we get
\be \label{FppTYw}
F_{(\p^p T\mY)}(w) = \f{c}{12} \sum_q \f{p!(q+3)!}{(p+q+4)!} F_{\mY,\p^q T}[f(w)]
                     \p_z^{p+q+4}\Big\{ \f{(z-w)^{q+4}f'(z)^2}{[f(z)-f(w)]^{q+4}} \Big\}_{z=w}
                     +O(s^2).
\ee
For an $\mathrm{SL}(2,\rC)$ conformal transformation $f(z)=\f{\a z+\b}{\g z+\d}$ with constants $\a,\b,\g,\d$ satisfying $\a \d-\b\g=1$,%
\footnote{One should not confuse the constant $\b$ here with the inverse temperature used in other places of the paper.}
we have
\be
\p_z^{p+q+4}\Big\{ \f{(z-w)^{q+4}f'(z)^2}{[f(z)-f(w)]^{q+4}} \Big\}_{z=w}
= \p_z^{p+q+4} [ (\g z+\d)^q ]_{z=w} (\g w+\d)^{q+4} = 0,
\ee
and so for a general conformal transformation $f(z)$ with a small $s(z)$ we have at least
\be
\p_z^{p+q+4}\Big[ \f{(z-w)^{q+4}f'(z)^2}{(f(z)-f(w))^{q+4}} \Big]_{z=w} = O(s).
\ee
For quasiprimary operator $\mY \neq T$, we have at least $F_{\mY,\p^q T} = O(s)$, and so we get
\be \label{put1} F_{(\p^p T\mY)} = O(s^2). \ee

For $\mY=T$, we get from (\ref{FppTYw})
\be
F_{(\p^p TT)}(w) = \f{c p!}{2(p+4)!} \p_z^{p+4}\Big[ \f{(z-w)^{4}f'(z)^2f'(w)^2}{(f(z)-f(w))^4} \Big]_{z=w} + O(s^2).
\ee
To evaluate it we need the Aharonov invariants $\psi_p$ that are defined as \cite{Aharonov:1969}
\be
\f{(z-w)^{2}f'(z)f'(w)}{[f(z)-f(w)]^2} = 1 + \sum_{p=2}^{+\inf} (p-1)(z-w)^p \psi_p(w).
\ee
Note that $\psi_2 = \f{s}{6}$. For $p\geq 3$, there is the nonlinear recursive formula \cite{Aharonov:1969}
\be
\psi_p = \f{1}{p+1} \Big( \psi_{p-1}' + \sum_{q=2}^{p-2}\psi_q\psi_{p-q} \Big).
\ee
We get for $p\geq2$
\be
\psi_p = \f{1}{(p+1)!}s^{(p-2)} + O(s^2).
\ee
Then we obtain
\be \label{put2}
F_{(\p^p TT)} = \f{c}{(p+1)(p+2)(p+4)(p+5)} s^{(p+2)} + O(s^2).
\ee

Note that
\be \label{put3}
F_{\p^p T} = \f{c}{12}s^{(p)}.
\ee
From (\ref{put1}), (\ref{put2}), (\ref{put3}) and (\ref{in}), we get for $h_\mX\geq4$
\be
F_\mX = \sum_{\mY \neq T} v_{\mX\mY} F_\mY^{(h_\mX-h_\mY)} + O(s^2).
\ee
From $F_\mA=O(s^2)$, we get by induction $F_\mX=O(s^2)$ for all holomorphic quasiprimary operators in the vacuum conformal family with $h_\mX\geq4$. Thus we prove that $a_\mX=0$ for $h_\mX>2$.

\providecommand{\href}[2]{#2}\begingroup\raggedright\endgroup


\begin{thebibliography}{}

\bibitem{Deutsch:1991}
J.~M. Deutsch, \textit{Quantum statistical mechanics in a closed system},
  \href{http://dx.doi.org/10.1103/PhysRevA.43.2046}{\textit{Phys. Rev.}
  {\bfseries A43} (1991) 2046--2049}.

\bibitem{Srednicki:1994}
M.~Srednicki, \textit{Chaos and quantum thermalization},
  \href{http://dx.doi.org/10.1103/PhysRevE.50.888}{\textit{Phys. Rev.}
  {\bfseries E50} (1994) 888--901}.

\bibitem{Srednicki:1995pt}
M.~Srednicki, \textit{{Thermal fluctuations in quantized chaotic systems}},
  \href{http://dx.doi.org/10.1088/0305-4470/29/4/003}{\textit{J. Phys.}
  {\bfseries A29} (1996) L75--L79},
  [\href{https://arxiv.org/abs/chao-dyn/9511001}{{\ttfamily
  chao-dyn/9511001}}].

\bibitem{Lashkari:2016vgj}
N.~Lashkari, A.~Dymarsky and H.~Liu, \textit{{Eigenstate Thermalization
  Hypothesis in Conformal Field Theory}},
  \href{http://dx.doi.org/10.1088/1742-5468/aab020}{\textit{J. Stat. Mech.}
  {\bfseries 1803} (2018) 033101},
  [\href{https://arxiv.org/abs/1610.00302}{{\ttfamily 1610.00302}}].

\bibitem{Dymarsky:2016aqv}
A.~Dymarsky, N.~Lashkari and H.~Liu, \textit{{Subsystem ETH}},
  \href{http://dx.doi.org/10.1103/PhysRevE.97.012140}{\textit{Phys. Rev.}
  {\bfseries E97} (2018) 012140},
  [\href{https://arxiv.org/abs/1611.08764}{{\ttfamily 1611.08764}}].

\bibitem{Brown:1986nw}
J.~D. Brown and M.~Henneaux, \textit{{Central charges in the canonical
  realization of asymptotic symmetries: an example from three-dimensional
  gravity}}, \href{http://dx.doi.org/10.1007/BF01211590}{\textit{Commun. Math.
  Phys.} {\bfseries 104} (1986) 207--226}.

\bibitem{Maldacena:1997re}
J.~M. Maldacena, \textit{{The Large N limit of superconformal field theories
  and supergravity}},
  \href{http://dx.doi.org/10.4310/ATMP.1998.v2.n2.a1}{\textit{Adv. Theor. Math.
  Phys.} {\bfseries 2} (1998) 231--252},
  [\href{https://arxiv.org/abs/hep-th/9711200}{{\ttfamily hep-th/9711200}}].
  [\href{http://dx.doi.org/10.1023/A:1026654312961}{\textit{Int. J. Theor.
  Phys.} {\bfseries 38} (1999) 1113--1133}].

\bibitem{Gubser:1998bc}
S.~Gubser, I.~R. Klebanov and A.~M. Polyakov, \textit{{Gauge theory correlators
  from noncritical string theory}},
  \href{http://dx.doi.org/10.1016/S0370-2693(98)00377-3}{\textit{Phys. Lett.}
  {\bfseries B428} (1998) 105--114},
  [\href{https://arxiv.org/abs/hep-th/9802109}{{\ttfamily hep-th/9802109}}].

\bibitem{Witten:1998qj}
E.~Witten, \textit{{Anti-de Sitter space and holography}},
  \href{http://dx.doi.org/10.4310/ATMP.1998.v2.n2.a2}{\textit{Adv. Theor. Math.
  Phys.} {\bfseries 2} (1998) 253--291},
  [\href{https://arxiv.org/abs/hep-th/9802150}{{\ttfamily hep-th/9802150}}].

\bibitem{Banados:1992wn}
M.~Ba\~nados, C.~Teitelboim and J.~Zanelli, \textit{{The Black hole in
  three-dimensional space-time}},
  \href{http://dx.doi.org/10.1103/PhysRevLett.69.1849}{\textit{Phys. Rev.
  Lett.} {\bfseries 69} (1992) 1849--1851},
  [\href{https://arxiv.org/abs/hep-th/9204099}{{\ttfamily hep-th/9204099}}].

\bibitem{Kraus:2016nwo}
P.~Kraus and A.~Maloney, \textit{{A Cardy formula for three-point coefficients
  or how the black hole got its spots}},
  \href{http://dx.doi.org/10.1007/JHEP05(2017)160}{\textit{JHEP} {\bfseries
  1705} (2017) 160}, [\href{https://arxiv.org/abs/1608.03284}{{\ttfamily
  1608.03284}}].

\bibitem{Brehm:2018ipf}
E.~M. Brehm, D.~Das and S.~Datta, \textit{{Probing thermality beyond the
  diagonal}}, \href{http://dx.doi.org/10.1103/PhysRevD.98.126015}{\textit{Phys.
  Rev.} {\bfseries D98} (2018) 126015},
  [\href{https://arxiv.org/abs/1804.07924}{{\ttfamily 1804.07924}}].

\bibitem{Romero-Bermudez:2018dim}
A.~Romero-Berm\'udez, P.~Sabella-Garnier and K.~Schalm, \textit{{A Cardy
  formula for off-diagonal three-point coefficients; or, how the geometry
  behind the horizon gets disentangled}},
  \href{http://dx.doi.org/10.1007/JHEP09(2018)005}{\textit{JHEP} {\bfseries
  1809} (2018) 005}, [\href{https://arxiv.org/abs/1804.08899}{{\ttfamily
  1804.08899}}].

\bibitem{Hikida:2018khg}
Y.~Hikida, Y.~Kusuki and T.~Takayanagi, \textit{{Eigenstate thermalization
  hypothesis and modular invariance of two-dimensional conformal field
  theories}}, \href{http://dx.doi.org/10.1103/PhysRevD.98.026003}{\textit{Phys.
  Rev.} {\bfseries D98} (2018) 026003},
  [\href{https://arxiv.org/abs/1804.09658}{{\ttfamily 1804.09658}}].

\bibitem{Calabrese:2004eu}
P.~Calabrese and J.~L. Cardy, \textit{{Entanglement entropy and quantum field
  theory}},
  \href{http://dx.doi.org/10.1088/1742-5468/2004/06/P06002}{\textit{J. Stat.
  Mech.} {\bfseries 0406} (2004) P06002},
  [\href{https://arxiv.org/abs/hep-th/0405152}{{\ttfamily hep-th/0405152}}].

\bibitem{Headrick:2010zt}
M.~Headrick, \textit{{Entanglement R\'enyi entropies in holographic theories}},
  \href{http://dx.doi.org/10.1103/PhysRevD.82.126010}{\textit{Phys. Rev.}
  {\bfseries D82} (2010) 126010},
  [\href{https://arxiv.org/abs/1006.0047}{{\ttfamily 1006.0047}}].

\bibitem{Calabrese:2010he}
P.~Calabrese, J.~Cardy and E.~Tonni, \textit{{Entanglement entropy of two
  disjoint intervals in conformal field theory II}},
  \href{http://dx.doi.org/10.1088/1742-5468/2011/01/P01021}{\textit{J. Stat.
  Mech.} {\bfseries 1101} (2011) P01021},
  [\href{https://arxiv.org/abs/1011.5482}{{\ttfamily 1011.5482}}].

\bibitem{Chen:2013kpa}
B.~Chen and J.-j. Zhang, \textit{{On short interval expansion of R\'enyi
  entropy}}, \href{http://dx.doi.org/10.1007/JHEP11(2013)164}{\textit{JHEP}
  {\bfseries 1311} (2013) 164},
  [\href{https://arxiv.org/abs/1309.5453}{{\ttfamily 1309.5453}}].

\bibitem{Sasaki:1987mm}
R.~Sasaki and I.~Yamanaka, \textit{{Virasoro Algebra, Vertex Operators, Quantum
  {Sine-Gordon} and Solvable Quantum Field Theories}},
  \href{http://dx.doi.org/10.1016/B978-0-12-385340-0.50012-7}{\textit{Adv.
  Stud. Pure Math.} {\bfseries 16} (1988) 271--296}.

\bibitem{Eguchi:1989hs}
T.~Eguchi and S.-K. Yang, \textit{{Deformations of Conformal Field Theories and
  Soliton Equations}},
  \href{http://dx.doi.org/10.1016/0370-2693(89)91463-9}{\textit{Phys. Lett.}
  {\bfseries B224} (1989) 373--378}.

\bibitem{Bazhanov:1994ft}
V.~V. Bazhanov, S.~L. Lukyanov and A.~B. Zamolodchikov, \textit{{Integrable
  structure of conformal field theory, quantum KdV theory and thermodynamic
  Bethe ansatz}}, \href{http://dx.doi.org/10.1007/BF02101898}{\textit{Commun.
  Math. Phys.} {\bfseries 177} (1996) 381--398},
  [\href{https://arxiv.org/abs/hep-th/9412229}{{\ttfamily hep-th/9412229}}].

\bibitem{Fitzpatrick:2014vua}
A.~L. Fitzpatrick, J.~Kaplan and M.~T. Walters, \textit{{Universality of
  Long-Distance AdS Physics from the CFT Bootstrap}},
  \href{http://dx.doi.org/10.1007/JHEP08(2014)145}{\textit{JHEP} {\bfseries
  1408} (2014) 145}, [\href{https://arxiv.org/abs/1403.6829}{{\ttfamily
  1403.6829}}].

\bibitem{Fitzpatrick:2015zha}
A.~L. Fitzpatrick, J.~Kaplan and M.~T. Walters, \textit{{Virasoro Conformal
  Blocks and Thermality from Classical Background Fields}},
  \href{http://dx.doi.org/10.1007/JHEP11(2015)200}{\textit{JHEP} {\bfseries
  1511} (2015) 200}, [\href{https://arxiv.org/abs/1501.05315}{{\ttfamily
  1501.05315}}].

\bibitem{Asplund:2014coa}
C.~T. Asplund, A.~Bernamonti, F.~Galli and T.~Hartman, \textit{{Holographic
  Entanglement Entropy from 2d CFT: Heavy States and Local Quenches}},
  \href{http://dx.doi.org/10.1007/JHEP02(2015)171}{\textit{JHEP} {\bfseries
  1502} (2015) 171}, [\href{https://arxiv.org/abs/1410.1392}{{\ttfamily
  1410.1392}}].

\bibitem{Caputa:2014eta}
P.~Caputa, J.~Sim{\'o}n, A.~$\check{\rm S}$tikonas and T.~Takayanagi,
  \textit{{Quantum Entanglement of Localized Excited States at Finite
  Temperature}}, \href{http://dx.doi.org/10.1007/JHEP01(2015)102}{\textit{JHEP}
  {\bfseries 1501} (2015) 102},
  [\href{https://arxiv.org/abs/1410.2287}{{\ttfamily 1410.2287}}].

\bibitem{Lin:2016dxa}
F.-L. Lin, H.~Wang and J.-j. Zhang, \textit{{Thermality and excited state
  R\'enyi entropy in two-dimensional CFT}},
  \href{http://dx.doi.org/10.1007/JHEP11(2016)116}{\textit{JHEP} {\bfseries
  1611} (2016) 116}, [\href{https://arxiv.org/abs/1610.01362}{{\ttfamily
  1610.01362}}].

\bibitem{He:2017vyf}
S.~He, F.-L. Lin and J.-j. Zhang, \textit{{Subsystem eigenstate thermalization
  hypothesis for entanglement entropy in CFT}},
  \href{http://dx.doi.org/10.1007/JHEP08(2017)126}{\textit{JHEP} {\bfseries
  1708} (2017) 126}, [\href{https://arxiv.org/abs/1703.08724}{{\ttfamily
  1703.08724}}].

\bibitem{Basu:2017kzo}
P.~Basu, D.~Das, S.~Datta and S.~Pal, \textit{{Thermality of eigenstates in
  conformal field theories}},
  \href{http://dx.doi.org/10.1103/PhysRevE.96.022149}{\textit{Phys. Rev.}
  {\bfseries E96} (2017) 022149},
  [\href{https://arxiv.org/abs/1705.03001}{{\ttfamily 1705.03001}}].

\bibitem{He:2017txy}
S.~He, F.-L. Lin and J.-j. Zhang, \textit{{Dissimilarities of reduced density
  matrices and eigenstate thermalization hypothesis}},
  \href{http://dx.doi.org/10.1007/JHEP12(2017)073}{\textit{JHEP} {\bfseries
  1712} (2017) 073}, [\href{https://arxiv.org/abs/1708.05090}{{\ttfamily
  1708.05090}}].

\bibitem{Lashkari:2017hwq}
N.~Lashkari, A.~Dymarsky and H.~Liu, \textit{{Universality of Quantum
  Information in Chaotic CFTs}},
  \href{http://dx.doi.org/10.1007/JHEP03(2018)070}{\textit{JHEP} {\bfseries
  1803} (2018) 070}, [\href{https://arxiv.org/abs/1710.10458}{{\ttfamily
  1710.10458}}].

\bibitem{Rigol:2006}
M.~{Rigol}, V.~{Dunjko}, V.~{Yurovsky} and M.~{Olshanii}, \textit{{Relaxation
  in a Completely Integrable Many-Body Quantum System: An AbInitio Study of the
  Dynamics of the Highly Excited States of 1D Lattice Hard-Core Bosons}},
  \href{http://dx.doi.org/10.1103/PhysRevLett.98.050405}{\textit{Phys. Rev.
  Lett.} {\bfseries 98} (2007) 050405},
  [\href{https://arxiv.org/abs/cond-mat/0604476}{{\ttfamily
  cond-mat/0604476}}].

\bibitem{Hartman:2013mia}
T.~Hartman, \textit{{Entanglement Entropy at Large Central Charge}},
  \href{https://arxiv.org/abs/1303.6955}{{\ttfamily 1303.6955}}.

\bibitem{Wong:2013gua}
G.~Wong, I.~Klich, L.~A. Pando~Zayas and D.~Vaman, \textit{{Entanglement
  Temperature and Entanglement Entropy of Excited States}},
  \href{http://dx.doi.org/10.1007/JHEP12(2013)020}{\textit{JHEP} {\bfseries
  1312} (2013) 020}, [\href{https://arxiv.org/abs/1305.3291}{{\ttfamily
  1305.3291}}].

\bibitem{Guo:2018fnv}
W.-Z. Guo, F.-L. Lin and J.~Zhang, \textit{{Non-geometric States in a
  Holographic Conformal Field Theory}},
  \href{https://arxiv.org/abs/1806.07595}{{\ttfamily 1806.07595}}.

\bibitem{Ryu:2006bv}
S.~Ryu and T.~Takayanagi, \textit{{Holographic derivation of entanglement
  entropy from AdS/CFT}},
  \href{http://dx.doi.org/10.1103/PhysRevLett.96.181602}{\textit{Phys. Rev.
  Lett.} {\bfseries 96} (2006) 181602},
  [\href{https://arxiv.org/abs/hep-th/0603001}{{\ttfamily hep-th/0603001}}].

\bibitem{Hubeny:2007xt}
V.~E. Hubeny, M.~Rangamani and T.~Takayanagi, \textit{{A Covariant holographic
  entanglement entropy proposal}},
  \href{http://dx.doi.org/10.1088/1126-6708/2007/07/062}{\textit{JHEP}
  {\bfseries 0707} (2007) 062},
  [\href{https://arxiv.org/abs/0705.0016}{{\ttfamily 0705.0016}}].

\bibitem{Dong:2016fnf}
X.~Dong, \textit{{The gravity dual of R\'enyi entropy}},
  \href{http://dx.doi.org/10.1038/ncomms12472}{\textit{Nature Commun.}
  {\bfseries 7} (2016) 12472},
  [\href{https://arxiv.org/abs/1601.06788}{{\ttfamily 1601.06788}}].

\bibitem{Banados:1998gg}
M.~Banados, \textit{{Three-dimensional quantum geometry and black holes}},
  \href{http://dx.doi.org/10.1063/1.59661}{\textit{AIP Conf. Proc.} {\bfseries
  484} (1999) 147--169},
  [\href{https://arxiv.org/abs/hep-th/9901148}{{\ttfamily hep-th/9901148}}].

\bibitem{Guo:2018djz}
W.-Z. Guo, F.-L. Lin and J.~Zhang, \textit{{Distinguishing Black Hole
  Microstates using Holevo Information}},
  \href{http://dx.doi.org/10.1103/PhysRevLett.121.251603}{\textit{Phys. Rev.
  Lett.} {\bfseries 121} (2018) 251603},
  [\href{https://arxiv.org/abs/1808.02873}{{\ttfamily 1808.02873}}].

\bibitem{Cardy:1986ie}
J.~L. Cardy, \textit{{Operator content of two-dimensional conformally invariant
  theories}},
  \href{http://dx.doi.org/10.1016/0550-3213(86)90552-3}{\textit{Nucl. Phys.}
  {\bfseries B270} (1986) 186--204}.

\bibitem{Ginsparg:1988ui}
P.~H. Ginsparg, \textit{{Applied Conformal Field Theory}},  in \textit{{Les
  Houches Summer School in Theoretical Physics: Fields, Strings, Critical
  Phenomena Les Houches, France, June 28-August 5, 1988}}, pp.~1--168, 1988.
\newblock \href{https://arxiv.org/abs/hep-th/9108028}{{\ttfamily
  hep-th/9108028}}.

\bibitem{DiFrancesco:1997nk}
P.~Di~Francesco, P.~Mathieu and D.~S\'en\'echal, \textit{{Conformal Field
  Theory}}.
\newblock Springer, New York, USA, 1997,
  \href{http://dx.doi.org/10.1007/978-1-4612-2256-9}{10.1007/978-1-4612-2256-9}.

\bibitem{Blumenhagen:2009zz}
R.~Blumenhagen and E.~Plauschinn, \textit{{Introduction to conformal field
  theory}}, \href{http://dx.doi.org/10.1007/978-3-642-00450-6}{\textit{Lect.
  Notes Phys.} {\bfseries 779} (2009) 1--256}.

\bibitem{Chen:2013dxa}
B.~Chen, J.~Long and J.-j. Zhang, \textit{{Holographic R\'enyi entropy for CFT
  with $W$ symmetry}},
  \href{http://dx.doi.org/10.1007/JHEP04(2014)041}{\textit{JHEP} {\bfseries
  1404} (2014) 041}, [\href{https://arxiv.org/abs/1312.5510}{{\ttfamily
  1312.5510}}].

\bibitem{Li:2016pwu}
Z.~Li and J.-j. Zhang, \textit{{On one-loop entanglement entropy of two short
  intervals from OPE of twist operators}},
  \href{http://dx.doi.org/10.1007/JHEP05(2016)130}{\textit{JHEP} {\bfseries
  1605} (2016) 130}, [\href{https://arxiv.org/abs/1604.02779}{{\ttfamily
  1604.02779}}].

\bibitem{Chen:2016lbu}
B.~Chen, J.-B. Wu and J.-j. Zhang, \textit{{Short interval expansion of R\'enyi
  entropy on torus}},
  \href{http://dx.doi.org/10.1007/JHEP08(2016)130}{\textit{JHEP} {\bfseries
  1608} (2016) 130}, [\href{https://arxiv.org/abs/1606.05444}{{\ttfamily
  1606.05444}}].

\bibitem{Cardy:2007mb}
J.~L. Cardy, O.~A. Castro-Alvaredo and B.~Doyon, \textit{{Form factors of
  branch-point twist fields in quantum integrable models and entanglement
  entropy}}, \href{http://dx.doi.org/10.1007/s10955-007-9422-x}{\textit{J.
  Stat. Phys.} {\bfseries 130} (2008) 129--168},
  [\href{https://arxiv.org/abs/0706.3384}{{\ttfamily 0706.3384}}].

\bibitem{Lashkari:2014yva}
N.~Lashkari, \textit{{Relative Entropies in Conformal Field Theory}},
  \href{http://dx.doi.org/10.1103/PhysRevLett.113.051602}{\textit{Phys. Rev.
  Lett.} {\bfseries 113} (2014) 051602},
  [\href{https://arxiv.org/abs/1404.3216}{{\ttfamily 1404.3216}}].

\bibitem{Lashkari:2015dia}
N.~Lashkari, \textit{{Modular Hamiltonian for Excited States in Conformal Field
  Theory}},
  \href{http://dx.doi.org/10.1103/PhysRevLett.117.041601}{\textit{Phys. Rev.
  Lett.} {\bfseries 117} (2016) 041601},
  [\href{https://arxiv.org/abs/1508.03506}{{\ttfamily 1508.03506}}].

\bibitem{Chen:2017ahf}
B.~Chen, Z.~Li and J.-j. Zhang, \textit{{Corrections to holographic
  entanglement plateau}},
  \href{http://dx.doi.org/10.1007/JHEP09(2017)151}{\textit{JHEP} {\bfseries
  1709} (2017) 151}, [\href{https://arxiv.org/abs/1707.07354}{{\ttfamily
  1707.07354}}].

\bibitem{Cardy:2014jwa}
J.~Cardy and C.~P. Herzog, \textit{{Universal Thermal Corrections to Single
  Interval Entanglement Entropy for Two Dimensional Conformal Field Theories}},
  \href{http://dx.doi.org/10.1103/PhysRevLett.112.171603}{\textit{Phys. Rev.
  Lett.} {\bfseries 112} (2014) 171603},
  [\href{https://arxiv.org/abs/1403.0578}{{\ttfamily 1403.0578}}].

\bibitem{Chen:2014unl}
B.~Chen and J.-q. Wu, \textit{{Single interval R\'enyi entropy at low
  temperature}}, \href{http://dx.doi.org/10.1007/JHEP08(2014)032}{\textit{JHEP}
  {\bfseries 1408} (2014) 032},
  [\href{https://arxiv.org/abs/1405.6254}{{\ttfamily 1405.6254}}].

\bibitem{Chen:2015uia}
B.~Chen, J.-q. Wu and Z.-c. Zheng, \textit{{Holographic R\'enyi entropy of
  single interval on torus: with W symmetry}},
  \href{http://dx.doi.org/10.1103/PhysRevD.92.066002}{\textit{Phys. Rev.}
  {\bfseries D92} (2015) 066002},
  [\href{https://arxiv.org/abs/1507.00183}{{\ttfamily 1507.00183}}].

\bibitem{Sarosi:2016oks}
G.~S\'arosi and T.~Ugajin, \textit{{Relative entropy of excited states in two
  dimensional conformal field theories}},
  \href{http://dx.doi.org/10.1007/JHEP07(2016)114}{\textit{JHEP} {\bfseries
  1607} (2016) 114}, [\href{https://arxiv.org/abs/1603.03057}{{\ttfamily
  1603.03057}}].

\bibitem{Sarosi:2016atx}
G.~S\'arosi and T.~Ugajin, \textit{{Relative entropy of excited states in
  conformal field theories of arbitrary dimensions}},
  \href{http://dx.doi.org/10.1007/JHEP02(2017)060}{\textit{JHEP} {\bfseries
  1702} (2017) 060}, [\href{https://arxiv.org/abs/1611.02959}{{\ttfamily
  1611.02959}}].

\bibitem{Aharonov:1969}
D.~Aharonov, \textit{A necessary and sufficient condition for univalence of a
  meromorphic function},
  \href{http://dx.doi.org/10.1215/S0012-7094-69-03671-0}{\textit{Duke Math. J.}
  {\bfseries 36} (1969) 599--604}.

\end{thebibliography}

\end{document}